\documentclass[aps, prd, twocolumn, groupedaddress,showpacs, nofootinbib]{revtex4-1}

\usepackage[english]{babel}
\usepackage[T1]{fontenc}
\usepackage[utf8]{inputenc}

\usepackage{lipsum}
\usepackage{amsmath}
\usepackage{graphicx}
\usepackage{amssymb}
\usepackage{multirow}
\usepackage{subfigure}
\usepackage{xspace}
\usepackage{enumerate}
\usepackage[dvipsnames]{xcolor}
\usepackage[normalem]{ulem}
\usepackage{mathtools}

\newcommand{\lsca}{\lambda_{\mathrm{scatt}}}
\newcommand{\lcoh}{\lambda_{\mathrm{coh}}}
\newcommand{\Rmax}{R_{\mathrm{max}}}
\newcommand{\Dmin}{D_{\mathrm{min}}}
\newcommand{\tmax}{t_{\mathrm{max}}}
\newcommand{\thub}{t_{\mathrm{Hub}}}
\newcommand{\ntot}{n_{\mathrm{tot}}}
\newcommand{\ares}{\theta_{\mathrm{res}}}
\newcommand{\fnd}{f_{\mathrm{non-dip}}}

\newcommand{\desy}{{DESY, D-15738 Zeuthen, Germany}}

\newcommand{\ifsc}{{Instituto de Física de São Carlos, Universidade de São Paulo, \\ Av. Trabalhador São-Carlense, 400, São Carlos, SP, Brazil}}

\begin{document}

\title{Ultrahigh-energy dipole and beyond}

\author{Rodrigo~Guedes~Lang}
\email{rodrigo.lang@usp.br}
\affiliation{\ifsc}

\author{Andrew~M.~Taylor}
\affiliation{\desy}

\author{Vitor~de~Souza}
\affiliation{\ifsc}

\date{\today}

\begin{abstract}
Recent data from the Pierre Auger Observatory has revealed the presence of a large-scale dipole in the arrival direction distribution of ultra-high energy cosmic rays (UHECR). In this work, we build up an understanding of the diffusive origin of such a dipolar behavior as well as its dependency on energy and astrophysical source assumptions such as extragalactic magnetic field strength and cosmic ray composition. We present a novel analytical approach for calculating the angular distribution of CR coming from a single source and discuss the regimes in which the steady-state dipole result is expected. We also present a semianalytical method for calculating the evolution with energy of the resultant dipole for an ensemble of sources. We show that a local source allows for a strong growth of the dipole with energy over a large energy range. The possibility of a transition from a dipolar to non-dipolar regime at the highest energies and its implications for the source density, magnetic field intensity, and cosmic ray composition are discussed.
\end{abstract}

\maketitle

\section{Introduction}
\label{sec:introduction}

One of the key unsolved questions in astroparticle physics is the origin of ultra-high energy cosmic rays (UHECR, $E>10^{18}$~eV)~\cite{AlvesBatista:2019tlv}. Discovered many decades ago, these nuclei reach Earth almost isotropically. The direction for their sources, however, is hidden by the deflections in the extra-galactic and galactic magnetic fields during their propagation. Results from the Pierre Auger Observatory~\cite{PierreAuger} using the most complete UHECR dataset ever collected have revealed the presence of a large-scale dipole in the distribution of arrival directions at energies above 8~EeV. Such  a dipole, with a magnitude of 6.5\%, was measured with 5.2$\sigma$ confidence level~\cite{AugerDipole} and points in a direction outwards away from the galactic center, suggestive of an extragalactic origin for cosmic rays at these energies. A more recent analysis from the Pierre Auger collaboration has furthermore revealed that the measured dipole magnitude itself evolves as a function of the energy, suggesting a change in the origin of the dipolar anisotropies from predominantly galactic to predominantly extra-galactic in the region between 1 and a few EeV~\cite{AugerRightAscension}.

In a previous work, we have discussed the possibility of using the arriving energy spectrum measured by the Pierre Auger Observatory~\cite{AugerSpectrum,AugerSpectrum1,AugerSpectrum2} to get insights about the radial distribution of UHECR sources~\cite{LangPRD2020}. UHECR interact with the photon background during their propagation, giving rise to an energy-dependent propagation horizon. From this, we estimated the maximum distance to the nearest source which still describes well the data, indicating the need for local sources at 25-100~Mpc.

In this work we further delve into the question about the origin of UHECR by investigating on theoretical grounds the evolution of the dipole magnitude as a function of energy for an ensemble of sources, and evaluating the influence of local sources on this result. Initially, in section~\ref{sec:singlesource}, we develop an analytical approach for calculating the resulting angular distribution of arriving cosmic rays propagating from a single source through an extragalactic environment within which turbulent magnetic fields are present. We obtain the amplitude of the dipole and higher poles by expanding this analytical distribution in spherical harmonics. The dependency with the magnetic field strength, the source distance, and the age of activity of the sources are evaluated and discussed. In section~\ref{sec:ensemble}, we use the results for a single source to obtain the evolution of the dipole with the energy expected from an ensemble of sources. In section~\ref{sec:results}, the results for different source densities, and consequently distances to the nearest source, are discussed. The transition from a dipolar to a non-dipolar regime is presented for the first time. Finally, in section~\ref{sec:conclusions}, we draw our conclusions.

\section{Cosmic Ray Angular Distribution From a Single Source}
\label{sec:singlesource}

Cosmic rays propagating through the extragalactic medium are deflected by the extragalactic magnetic fields. Such propagation can be considered as each particle having its direction of motion randomized each scattering length, $\lsca$. The scattering length is dependent on the Larmor radius, $R_L$, of the particle, and the level of magnetic turbulence at a wavelength matching the Larmor length scale.
Appealing to the results of quasi-linear theory~\cite{1989ApJ...336..243S}, $\lsca=[B_{0}^{2}/ \delta B(R_{L})^{2}]R_{L}$, where $\delta B(\lambda)^{2}= d \delta B ^{2}/d\ln \lambda$ denotes the differential energy density in logarithmic wavelength bins. With the energy density in the turbulent modes dominated by the longest wavelengths, the scattering length carries information about the ratio of energies in the coherent magnetic field, to that in the turbulent magnetic field, $B_{0}^{2}/\delta B^2$, where $\delta B^2 = \int \delta B(\lambda)^2 d\ln\lambda$ is the summed power of the turbulent modes. Leading to, $\delta B(R_{L})^2= (R_{L}/\lcoh)^{q-1}\delta B^2$, where $q$ is the spectral slope of turbulence ($5/3$ for Kolmogorov turbulence) and $R_{L}/\lcoh$ is the ratio of the particle's Larmor radius to the coherence length of the field. This allows a simple description of the scattering length, $\lsca$~\cite{OSullivan:2009rvg},

\begin{equation}
    \label{eq:lsca}
    \lsca = \left(\frac{B_{0}^{2}}{\delta B^{2}}\right)\lcoh\begin{cases}
    \left(\frac{R_{L}}{\lcoh} \right)^{1/3} , & \ \mathrm{for} \ R_L < \lcoh \\
    \left(\frac{R_{L}}{\lcoh} \right)^{2}, & \ \mathrm{for} \ R_L \ge \lcoh
    \end{cases}.
\end{equation}

No Galactic magnetic fields effects are here considered. Throughout we assume $\delta B ^{2}=B_{0}^{2}$.

Consequently, the particles arriving at Earth will not necessarily point back to their original source. Nevertheless, as we discuss next, some residual information about the source position invariably remains encoded in the arrival directions distribution.

\begin{figure}[ht]
    \centering
    \includegraphics[width=0.48\textwidth]{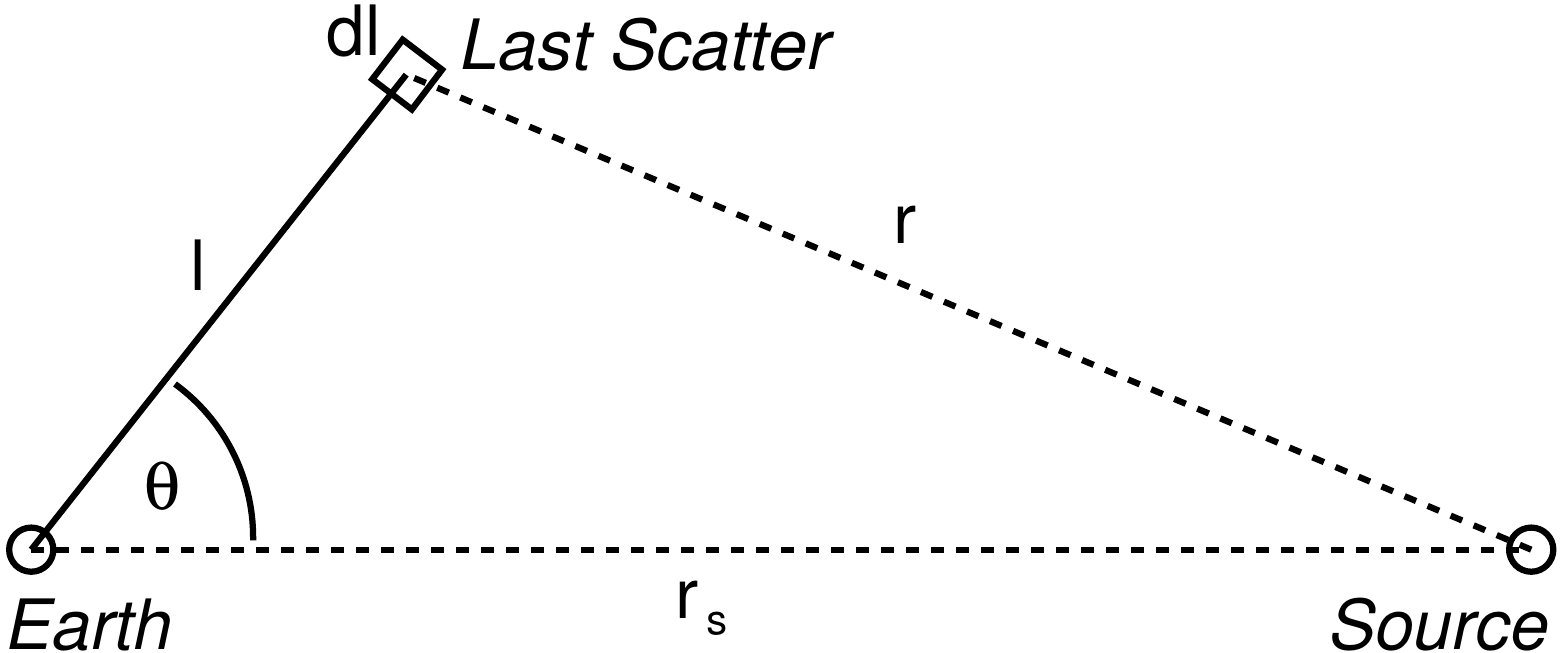}
    \caption{Diagram depicting the arrival of diffusively propagating particles from a source, focusing on the last scattering the particles make before arriving to Earth.}
    \label{fig:diagram}
\end{figure}

The angular distribution of CR coming from a single source was previously studied with Monte Carlo simulations in Ref.~\cite{Harari:2013pea}. As expected from diffusion theory, it was shown that, in the steady-state diffusive regime, achieved for sufficiently long source activity timescales, the normalized angular distribution can be described as $dN/d\cos \theta = 1 + \delta \cos \theta$,
where $\theta$ is the angle between the source position and the arrival direction (see fig.~\ref{fig:diagram}). In this regime, the dipole amplitude of cosmic rays of a given rigidity emanating from a single source relates to the particle scattering length and the source distance, $r_{s}$, as $\delta = \lsca/r_s$.

In this work, we have further developed this calculation and discuss the validity of the steady-state diffusive regime. We propose an analytical approach for obtaining the resultant CR angular distribution following the propagation of CR from their sources to Earth. To achieve this, we utilize earlier insights obtained for describing CR transport in different propagation regimes~\cite{LangPRD2020}. We here look into how these different propagation regimes imprint themselves onto the arriving CR angular distribution. Additionally, we test a variety of different source properties, such as the mean source distance and age of source activity.

We start by considering a single source, which emitting cosmic rays continuously and isotropically. Each of the particles has a probability of $dl/\lsca$ of having its direction of motion randomized after propagating a small distance $dl$. Previously, in Ref.~\cite{LangPRD2020}, we obtained the radial evolution distribution function for cosmic rays emitted in a single pulse, i.e., the radial Green's function from a single source, $d^{3}N/dr^{3}|_{\mathrm{G}} (r,t)$. Three regimes were noted. For $3ct/\lsca < 0.1$, the propagation is ballistic and a delta distribution can be used. For $3ct/\lsca > 10$, the propagation is diffusive and a truncated Gaussian can be used (truncated in order to prevent super luminal propagation). Finally, for the transition regime, $0.1 \le 3ct/\lsca \le 10$, a J\"uttner distribution can be used.

In order to obtain the angular distribution, we calculate the number of particles arriving to Earth at an angle $\theta$ relative the source direction, in a time interval, $dt$, and in a small area, $dA$, i.e., $d^2N/dAdt (\theta)$. The diagram in figure~\ref{fig:diagram} illustrates the geometry of the problem. As appreciated from this figure, for a particle to arrive to Earth with a given arrival direction, $\theta$, it must have last scattered somewhere in the line of sight along a path at that angle, and then subsequently have survived all the way to Earth without additional scattering. The number of particles that fulfill these criteria is given by

\begin{equation}
    \label{eq:diff}
    \left.\frac{d^2N}{dAdt}\right|_{\mathrm{diff}} (\theta, t) = \int_{0}^{\infty} \frac{dl e^{-l/\lsca} n \left(r(l), t+l\right)}{\lsca /c},
\end{equation}

where $c dt/\lsca$ is the fraction of particles that scattered in the time interval $dt$, $e^{-l/\lsca}$ is the probability that these particles survived all the way to Earth without interacting again, and $n (r, t+l)$ is the particle density at this location, given by

\begin{equation}
    n (r,t) = \int_{t}^{\tmax} \frac{dt'}{\tau} \left. \frac{d^3N}{dr^{3}} \right|_{\mathrm{G}} (r,t'),
\end{equation}

where $d^{3}N/dr^{3}|_{\mathrm{G}} (r,t)$ are the distributions obtained in Ref.~\cite{LangPRD2020}, $\tau^{-1}$ is the source emissivity taken as constant, $\tmax$ is the age of activity of the sources, and $r$ is the distance to the source, which relates to the distance to Earth, $l$, as

\begin{equation}
    \label{eq:r}
    r^2 = l^2 + r_s^2 - 2 l r_s \cos \theta.
\end{equation}

It is also necessary to take into the account the particles which arrived on Earth having propagated ballistically, i.e., without ever scattering, which is given by

\begin{equation}
    \label{eq:ballistic}
    \left.\frac{d^2N}{dAdt}\right|_{\mathrm{bal}} (\theta, t) = \delta (\theta) \frac{e^{-r_s/\lsca}}{\tau 4 \pi r_s^2},
\end{equation}

where $dt/\tau$ is the number of particle emitted in a time interval $dt$. The delta function assures that such particles can only arrive from the direction of the source, and $e^{-r_s/\lsca}$ gives the fraction of particles in this regime. Finally, the total number of particles arriving with a direction given by the angle $\theta$, in a time interval $dt$, and in a small area $dA$ is given by

\begin{equation}
    \label{eq:angulardist}
    \frac{d^2N}{dAdt} (\theta, t) = \left.\frac{d^2N}{dAdt}\right|_{\mathrm{diff}} (\theta, t) + \left.\frac{d^2N}{dAdt}\right|_{\mathrm{bal}} (\theta, t).
\end{equation}

At this stage, the angular distribution, $dN/d\cos \theta$, can be obtained by integrating $d^2N/dAdt$ over the exposure of the experiment. However, as the time scale of a human-made experiment is negligible compared to cosmological time scales, the factoring in of the exposure is effectively just a global normalization factor.

\begin{figure}[ht]
    \centering
    \includegraphics[width=0.49\textwidth]{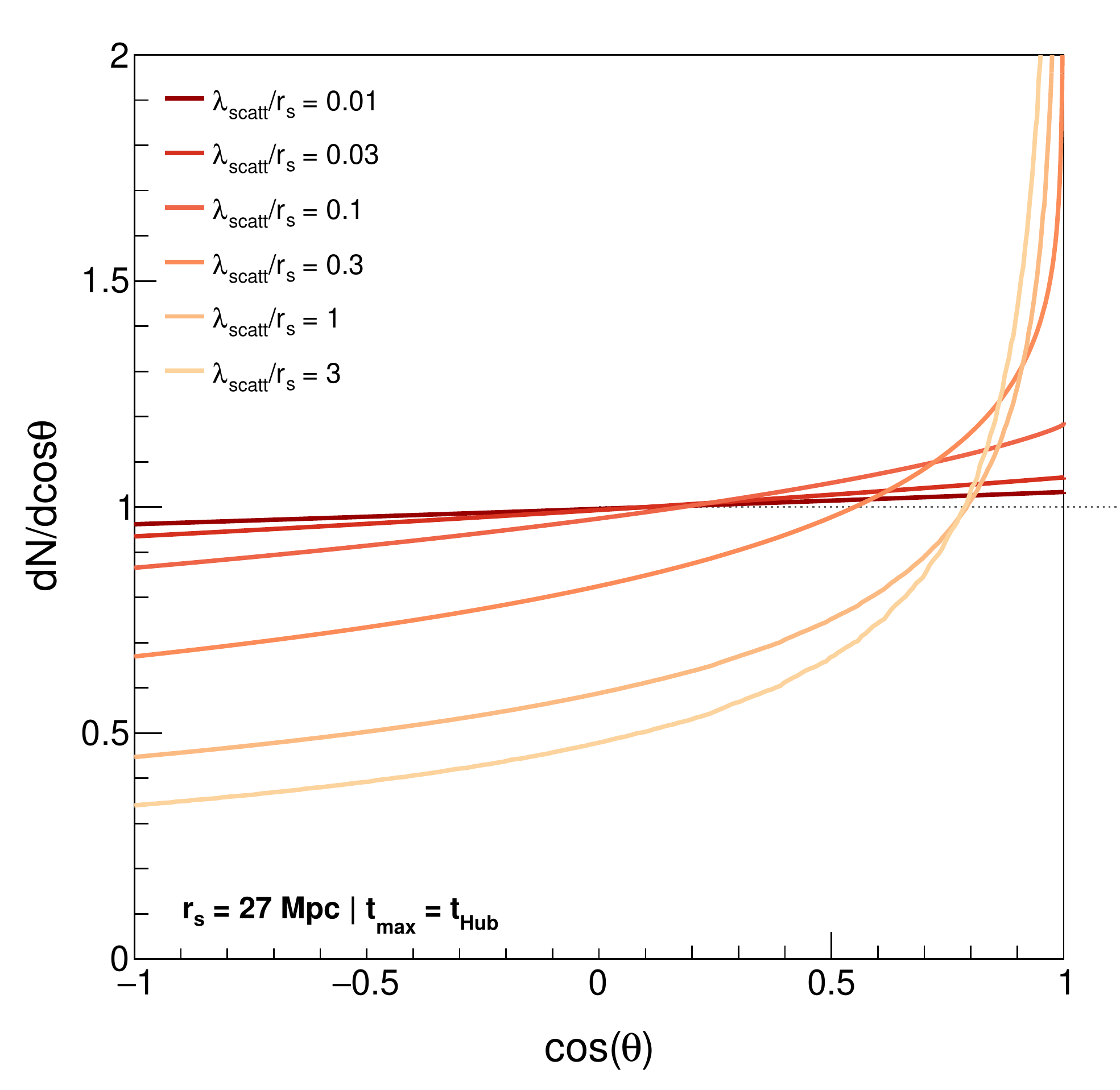}
    \caption{Normalized angular distribution. Each shade of orange represents a value of $\lsca/r_s$.}
    \label{fig:angdist}
\end{figure}

Solving this expression numerically, figure~\ref{fig:angdist} shows the resulting angular distribution for different values of $\lsca/r_s$. For large values, the distribution is dominated by the delta distribution coming from the ballistic propagation. For small values, on the other hand, the distribution is mostly uniform, with a small preference for the direction of the source, and can be well described by $1+\delta \cos \theta$.

\subsection{Steady-State Diffusive Regime}

For the steady-state diffusive regime ($\lsca/r_s > r_s/c\tmax$), the ballistic term vanishes and the density can be written as $n (r) \propto 1/(\lambda_{\rm scatt}r)$. The angular distribution is thus given by

\begin{equation}
    \begin{split}
    \frac{dN}{d\cos \theta} \propto \int_{0}^{\infty} \frac{dl e^{-l/\lsca}}{r(l)} \approx \frac{\lsca}{r(\lsca)} \\
    \approx \frac{\lsca}{r_s \left(1 - \frac{\lsca}{r_s} \cos \theta \right)} \approx \frac{\lsca}{r_s} \left(1 + \frac{\lsca}{r_s} \cos \theta\right).
    \end{split}
\end{equation}

Therefore, the results obtained in previous works and expected from the diffusive theory for the steady-state regime are verified.

\subsection{Expansion in Spherical Harmonics}

A more quantitative way of studying such distributions is to expand them in spherical harmonics and to look at the behavior of the coefficients. The angular distribution, $dN/d\cos \theta$, can be written in orthogonal functions,

\begin{equation}
    \frac{dN}{d\cos \theta} = N \left[1 +\sum_{n=1}^{\infty} \sum_{m=-\ell}^{m=\ell} \Phi_{\ell,m} Y_{\ell}^{m} (\phi, \theta)\right],
\end{equation}

For the case of a single source in turbulent magnetic fields, no dependency on $\phi$ is expected if $\theta$ is the angle between the source and the arrival direction as shown in figure~\ref{fig:diagram} and, consequently, only $m=0$ is needed. We define the spherical harmonics similarly to what is done in Ref.~\cite{Harari:2013pea},

\begin{equation}
    Y_{\ell} (\theta) = Y^{0}_{\ell} (\theta) = \frac{2\ell+1}{\sqrt{2}} P_{\ell} (\cos \theta),
\end{equation}

with $P_{\ell} (\cos \theta)$ being the Legendre polynomials. With that, the amplitude of the poles, $\Phi_{\ell}$, are calculated by

\begin{equation}
\label{eq:coefficients}
    \Phi_\ell = \int_{-1}^{1} d\cos \theta \frac{dN}{N d\cos \theta} \frac{2\ell+1}{\sqrt{2}} P_{\ell} (\cos \theta),
\end{equation}

in particular the amplitude of the dipole is defined as $\delta = \Phi_1$. We can obtain the information about the power in each multiple by performing an angular power spectrum analysis, similarly to Refs.~\cite{Ahlers:2016rox,Wittkowski:2017nfd,Aartsen:2018ppz},

\begin{equation}
    \label{eq:normalization}
    C_\ell = \frac{\Phi^2_\ell/(2\ell+1)}{\sum_n \Phi_{n}^2/(2n+1)},
\end{equation}

which leads to

\begin{equation}
\begin{split}
&0 < C_\ell < 1\mathrm{, \ for \ all} \ \ell \\
&\sum_{\ell=0}^{\infty} C_\ell = 1.
 \end{split}
\end{equation}

We also define the non-dipolarity of the distribution, $\fnd$, as the power in the multipoles higher than the dipole,

\begin{equation}
    \fnd = 1 - C_0 - C_1,
\end{equation}

which is negligible for dipolar distributions, i.e., distributions that are well described by $1+\delta \cos \theta$.

\subsection{Evolution of the coefficients of the spherical harmonics}

With these definitions, we can calculate the evolution of the amplitude of the poles, $\Phi_{\ell}$, and the power spectrum coefficients, $C_{\ell}$, with $\lsca/r_s$ for different source distance, $r_{s}$, and source activity duration, $t_{\rm max}$, cases. 

\begin{figure}[ht]
    \centering
    \includegraphics[width=0.49\textwidth]{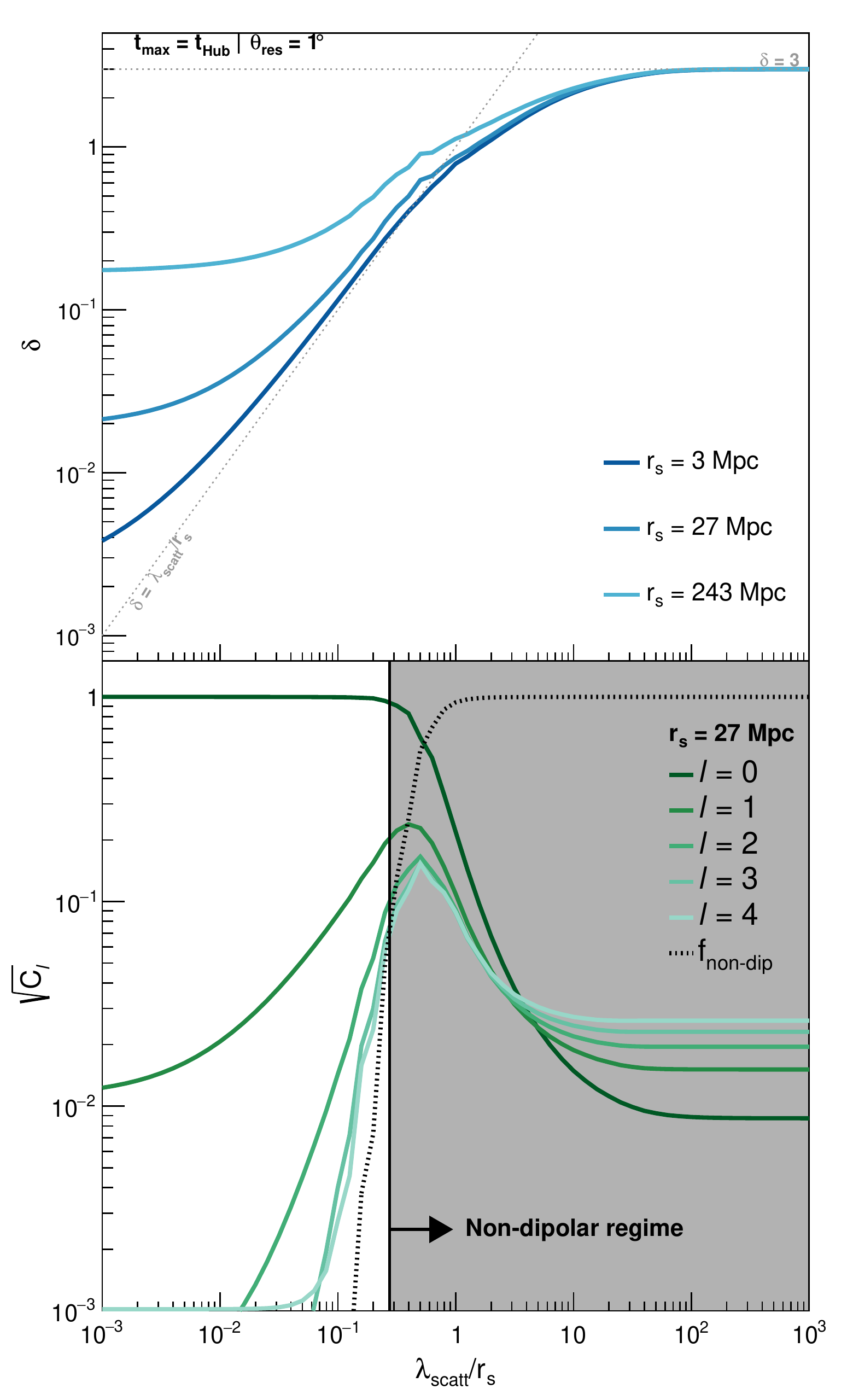}
    \caption{Evolution of the spherical harmonics expansion terms with $\lsca/r_s$. The top panel shows the amplitude of the dipole for different distances to source. The bottom panel shows the angular power spectrum for $r_s = 27$~Mpc. The gray dashed lines on the top panel show the results expected for the steady-state diffusive and ballistic regimes, while the black dashed line on the bottom panel shows the term $\fnd$. The shaded area shows the non-dipolar regime, in which $\fnd > 1\%$. $\thub$ is the Hubble time}
    \label{fig:singlesourcers}
\end{figure}

The top panel of figure~\ref{fig:singlesourcers} shows the values of $\delta$ obtained for different source distances. The dashed lines in the figure show the linear relation expected for transport in the steady-state diffusive regime, i.e., $\delta = \lsca/r_s$ and in the ballistic regime, i.e., $\delta = 3$. For values of $\lsca/r_s$ between $\sim r_s/c\tmax$ and $\sim$0.1, agreement with the linear relation result is obtained. In this regime, $\sqrt{3 C_1} \sim \Phi_1 = \delta$. However, outside this range a strong departure from this relation is seen.

For sufficiently low values of $\lsca/r_s$ ($\lambda_{\rm scatt.}/r_{s} < r_{s}/ct_{\rm max}$), the dependency of $\delta$ becomes softer than the steady-state result, since the system has insufficient time to reach this state, and the resulting dipole is larger than the steady-state result case. For a given $t_{\rm max}$, the farther the source, the sooner (in $\lambda_{\rm scatt.}/r_{s}$) that this departure kicks in. This is due to the finite age of the universe, which prevents the density of cosmic rays emitted from the source reaching the steady-state value. 

For sufficiently high values of $\lsca /r_{s}$ ($\lsca /r_{s} > 0.3$), on the other hand, the ballistic regime contribution becomes significant. This is measured by the angular power spectrum shown in the bottom panel of figure~\ref{fig:singlesourcers}. When the power in the higher order poles ($\ell>1$) becomes non-negligible, the dipole amplitude departures from the linear relation. In this regime, most of the particles do not diffuse at all and, therefore, point directly back to the source. This results in a non-dipolar behavior, which is verified by the increasing of the $\fnd$ term.\footnote{Throughout we do not consider small angles deflections, which would alter the transition from diffusive to ballistic and, thus, the growth of higher order multipoles.} $C_0$ and $C_1$ tend to constant small values which follows $C_0 = C_1/3$, as expected from the definition in equation~\ref{eq:coefficients}. This is due to a finite angular resolution.

The effects of the angular resolution, $\theta_{\mathrm{res}}$ and time of activity of the sources, are detailed in appendix~\ref{app:singlesource}.

\section{Dipole of an ensemble of sources}
\label{sec:ensemble}

In the previous section, we built an understanding of the angular distribution resulting from a single source of cosmic rays. In this section, we evaluate the evolution with energy of the dipole for an ensemble of sources. This was previously studied in Refs.~\cite{Harari:2015hba,Globus:2017fym,Dundovic:2017vsz,Mollerach:2020mhr}. We consider a discrete distribution of sources described by the distance to the nearest source, $\Dmin$, in such a way that the number of sources in a shell with distance $i\Dmin$ is $i^2$.

\subsection{Coefficients of the spherical harmonics for an ensemble of sources}

As proposed in section~\ref{sec:singlesource}, given the source distance, the magnetic field and the particle energy and charge, it is possible to analytically obtain the arriving angular distribution of cosmic rays for that source, $dN_i/d\cos \theta$, and consequently its harmonic coefficients, $\Phi_\ell^{(s)}$, where $(s)$ denotes the source. In this section, we discuss the approach for obtaining a total arriving angular distribution, $dN_{\mathrm{tot}}/d\cos \theta$.

Given two angular distributions, with two preferred angles, $dN_1/d\cos \theta_1$ and $dN_2/d\cos \theta_2$, one can obtain $dN_{\rm sum}/d\cos \theta_{\rm sum} = dN_1/d\cos \theta_1 + dN_2/d\cos \theta_2$, with $b_\ell^{(\rm sum)}(\theta_{\rm sum}) = 0$, for every $\ell$. Its harmonic coefficients will be given by

\begin{equation}
\label{eq:cos}
\begin{split}
\Phi_0^{(\rm sum)} &= \Phi_0^{(1)} + \Phi_0^{(2)} \\
\Phi_\ell^{(\rm sum)} &= \sqrt{\left(\Phi_\ell^{(1)}\right)^2 + \left(\Phi_\ell^{(2)}\right)^2 + 2 \Phi_\ell^{(1)} \Phi_\ell^{(2)} \cos(\alpha) },
\end{split}
\end{equation}

where $\alpha$ is the angle between $\theta_1$ and $\theta_2$. If we consider sources randomly distributed, then $\langle\cos(n\alpha)\rangle = 0$. Therefore, the coefficients of the angular distribution from an ensemble of sources will be given by

\begin{equation}
\label{eq:coefensemble}
\begin{split}
    \Phi_0^{\mathrm{(tot)}} (E) &= \sum_s n_s (E, r_s) \\
    \Phi_{\ell>0}^{\mathrm{(tot)}} (E) &= \sqrt{\sum_s \left[\Phi_\ell (E, r_s) n_s (E, r_s)\right]^2},
    \end{split}
\end{equation}

where $n_s$ denotes the arriving cosmic ray density, and the sum over $s$ denotes the sum over all sources. The angular power spectrum, which contain the information about the dipolarity of the distribution, can then be obtained by

\begin{equation}
\label{eq:coefensemble2}
    C_\ell^{\mathrm{(tot)}} (E) = \frac{\left(\Phi^{\mathrm{(tot)}}_\ell (E)\right)^{2} / (2\ell+1)}{\sum_{n=0}^{\infty} \left(\Phi_n^{\mathrm{(tot)}} (E)\right)^2/(2n+1)}.
\end{equation}

\subsection{The nearest source dipole and dilution from further sources}
\label{sec:shells}

During propagation, cosmic rays may interact with the photon background via pair production, pion production and photodisintegration, which leads to an energy-dependent propagation horizon~\cite{Hooper:2006tn,Allard:2011aa,Lemoine_2005,Aloisio_2005,Globus_2007}. Limitations on the propagation distance, which depends on the particle energy, are also expected due to the presence of extragalactic magnetic fields~\cite{Taylor11,Mollerach_2013,LangPRD2020}. We use the semianalytical method proposed in Ref.~\cite{LangPRD2020} to calculate the contribution from each source to the spectrum, $n_i (E, r_i)$.

\begin{figure}[ht]
    \centering
    \includegraphics[width=0.49\textwidth]{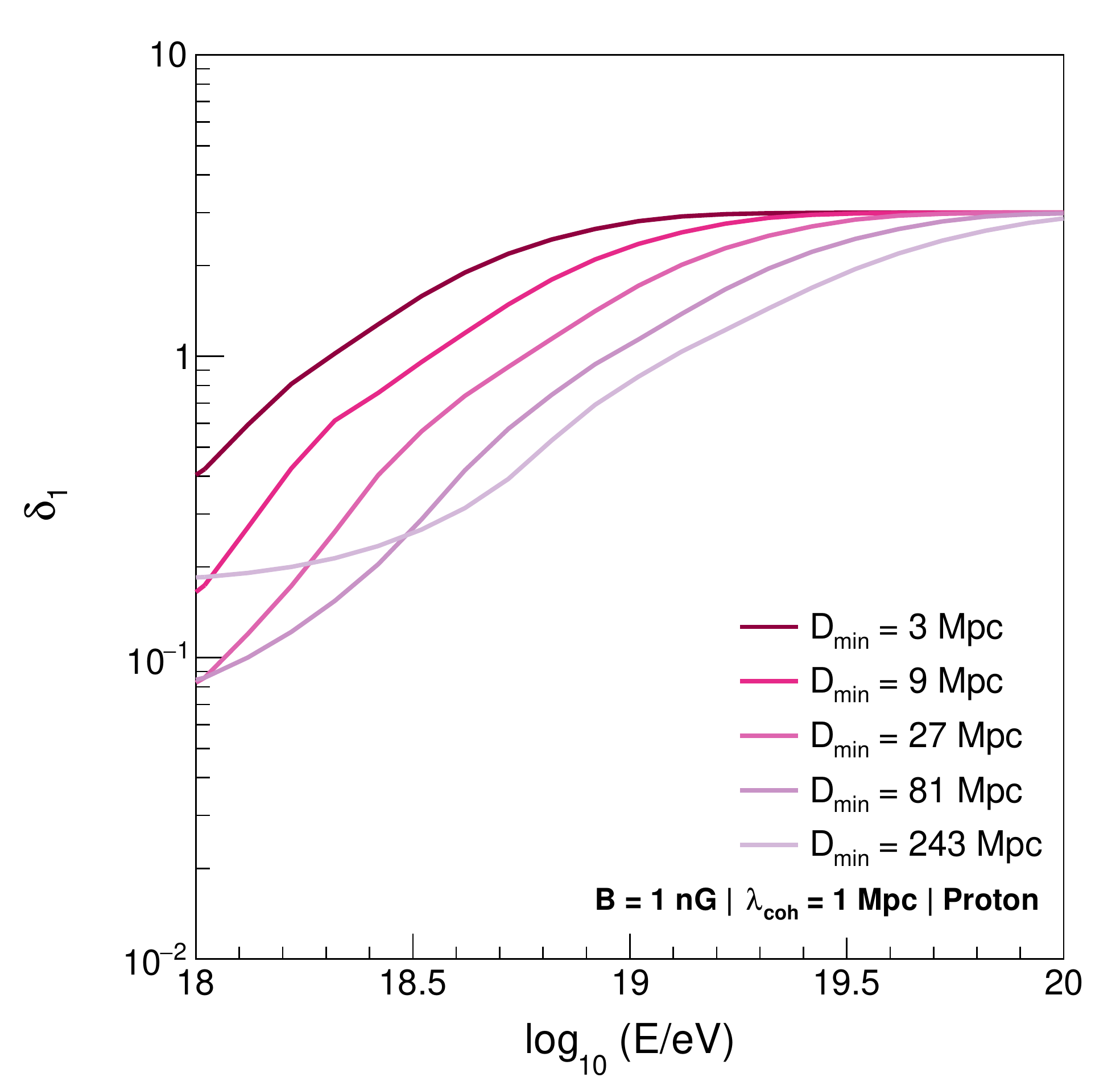}
    \caption{Dipole from the closest source as a function of the energy. Each colored line represents a different distance to the nearest source. A proton primary is considered, but similar results are found for the other primaries.}
    \label{fig:firstdipole}
\end{figure}

\begin{figure}[ht]
    \centering
    \includegraphics[width=0.49\textwidth]{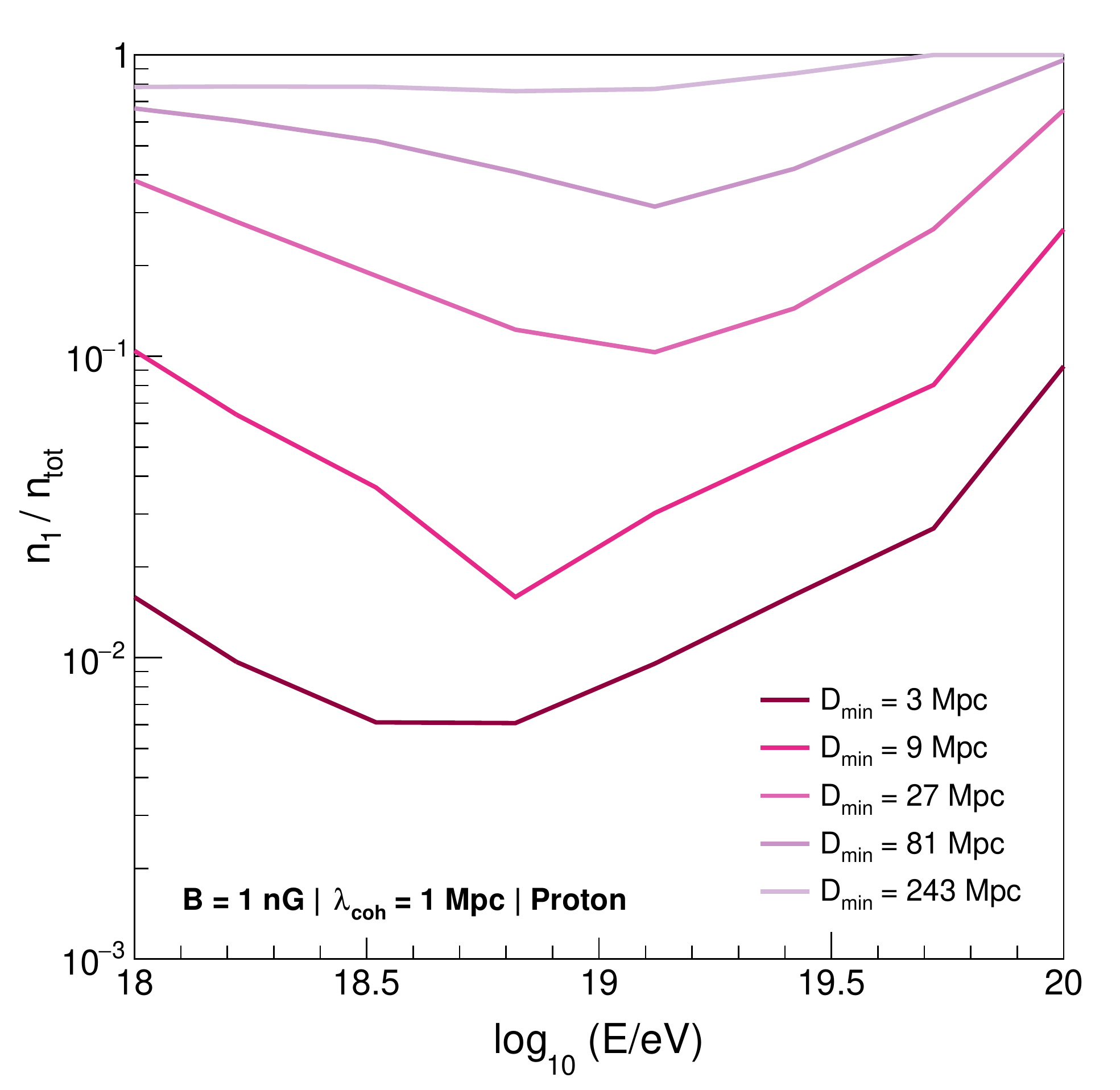}
    \caption{Dilution factor of the dipole from the nearest source due to the contribution of the more distant sources. Each colored line represents a different distance to the nearest source. A proton primary is considered, but similar results are found for the other primaries.}
    \label{fig:dilutionfactor}
\end{figure}

An understanding for the resultant dipole from an ensemble of sources is obtained from the consideration of equation~\ref{eq:coefensemble}. Specifically, it should be noted that the dipole term sums incoherently, whilst the isotropic term sums coherently. As further developed in appendix~\ref{app:approximation}, to a first approximation the total dipole can be understood as the dipole from the closest source diluted by the ratio of the contribution of farther sources, i.e., $\delta \approx \delta_1 n_1/n_{\rm tot}$.

Figures~\ref{fig:firstdipole} and \ref{fig:dilutionfactor} show, respectively, the dipole for the closest source ($\delta_1$) and the dilution factor ($n_{1}/n_{\rm tot}$), i.e., the ratio between the cosmic ray density coming from the closest source and total cosmic ray density. Figure~\ref{fig:firstdipole} can be appreciated as a re-expression of the relation shown in figure~\ref{fig:singlesourcers} in terms of energy.

Figure~\ref{fig:dilutionfactor} shows the dilution factor level due to the contribution from more distant sources. For both lower and higher energies, this dilution factor is smaller. The growth of the dilution factor at lower energies is brought about by the magnetic horizon effect, which suppresses the cosmic ray contribution from more distant source \cite{LangPRD2020}. At higher energies, the role of the energy loss horizon sets in, which suppresses the cosmic rays from farther sources.

\section{Evolution of the dipole with energy}
\label{sec:results}

With a clear understanding of the origin of an UHECR dipole, how the ensemble dipole evolves as a function of energy can be addressed. To obtain the dipole strength at each energy, the contribution of each source to both the spectrum and coefficients of the spherical harmonics must be obtained.

Following equation~\ref{eq:coefensemble} and the semi-analytical method to obtain the arriving cosmic ray rate considering energy losses and magnetic fields effects, we have calculated the evolution of the harmonic coefficients with energy for different ensembles. A spectral index of $\Gamma = 2$, a maximum rigidity of $\Rmax = 10^{19}$~V and pure composition at the sources were considered. Several distances to the nearest source, $\Dmin$, primary species, and magnetic field intensities, $B$, within the range set by observations~\cite{Kronberg_1976,Kronberg:1993vk,Blasi_1999,Schleicher:2011jj} were treated. 

\begin{figure}
    \centering
    \includegraphics[width=0.49\textwidth]{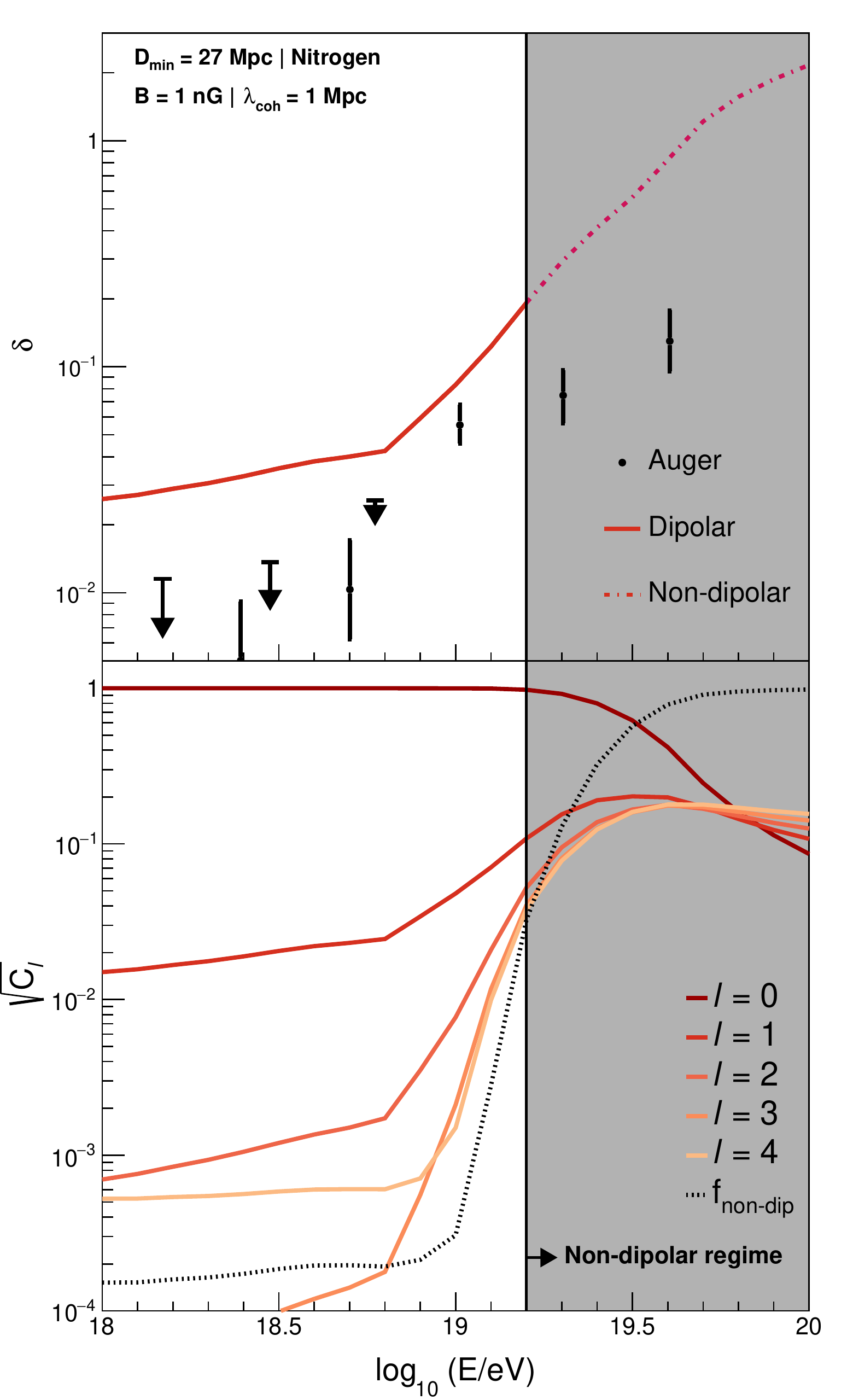}
    \caption{Evolution of the total angular distribution with energy. The top panel shows the amplitude of the dipole calculated from the model and measured by the Pierre Auger Observatory~\cite{AugerDipole}. The bottom panel shows the angular power spectrum. The shaded area represents the energy range on the non-dipolar regime.}
    \label{fig:powerspectrum}
\end{figure}

The top panel of figure~\ref{fig:powerspectrum} shows the resulting dipole, $\delta$, for an example case with $r_s = 27$~Mpc, $B = 1$~nG and nitrogen composition together the data measured the Pierre Auger Observatory~\cite{AugerDipole}. The bottom panel shows the evolution of the power spectrum with energy. Two main regimes are found. For lower energies, the total angular distribution is dipolar, i.e., most of the power of the function is in the first two poles ($\ell = 0$ and $\ell = 1$). For this regime, the normalized distribution is well described by $1 + \delta \cos \theta$. For higher energies, the non-dipolarity of the distribution increases and the power is divided in higher order multipoles. The transition between the regimes is quantified by $\fnd$ and we define the source to be in the dipolar regime when $\fnd < 1\%$.

The analysis done by the Pierre Auger Collaboration shows that the measured distribution of arrival directions can be well described by just a dipolar expansion and the data is consistent with the assumption that no multipoles of higher order are present. Henceforth, the angular power spectrum in not be explicitly shown and the amplitude of $\delta$ for the astrophysical models will be represented by continuous and dashed lines. The change from continuous to dashed lines is to illustrate the energy at which the non-dipolar behaviour becomes important for each astrophysical model, i.e., $\fnd > 1\%$. Continuous lines show the values of $\delta$ for energies at which $\fnd < 1\%$ for each astrophysical model. Dashed lines, on the other hand, show the values of $\delta$ for energies at which $\fnd > 1\%$.

A statistical comparison between the models and the data is beyond the scope of this work. However, it is important to note that a comparison of the models to the data would only be valid in the energy range delimited by the solid lines. In this energy range, both the data and the models can be well described by a pure dipolar expansion ($\fnd < 1\%$)  and, therefore, the comparison is consistent. A comparison of the value of $\delta$ obtained by the models to the data in the energy range shown by the dashed lines is not valid because in this range the data can be described by a pure dipolar expansion as shown in the analysis of the Pierre Auger Collaboration while the models cannot.

\begin{figure}[ht]
    \centering
    \includegraphics[width=0.49\textwidth]{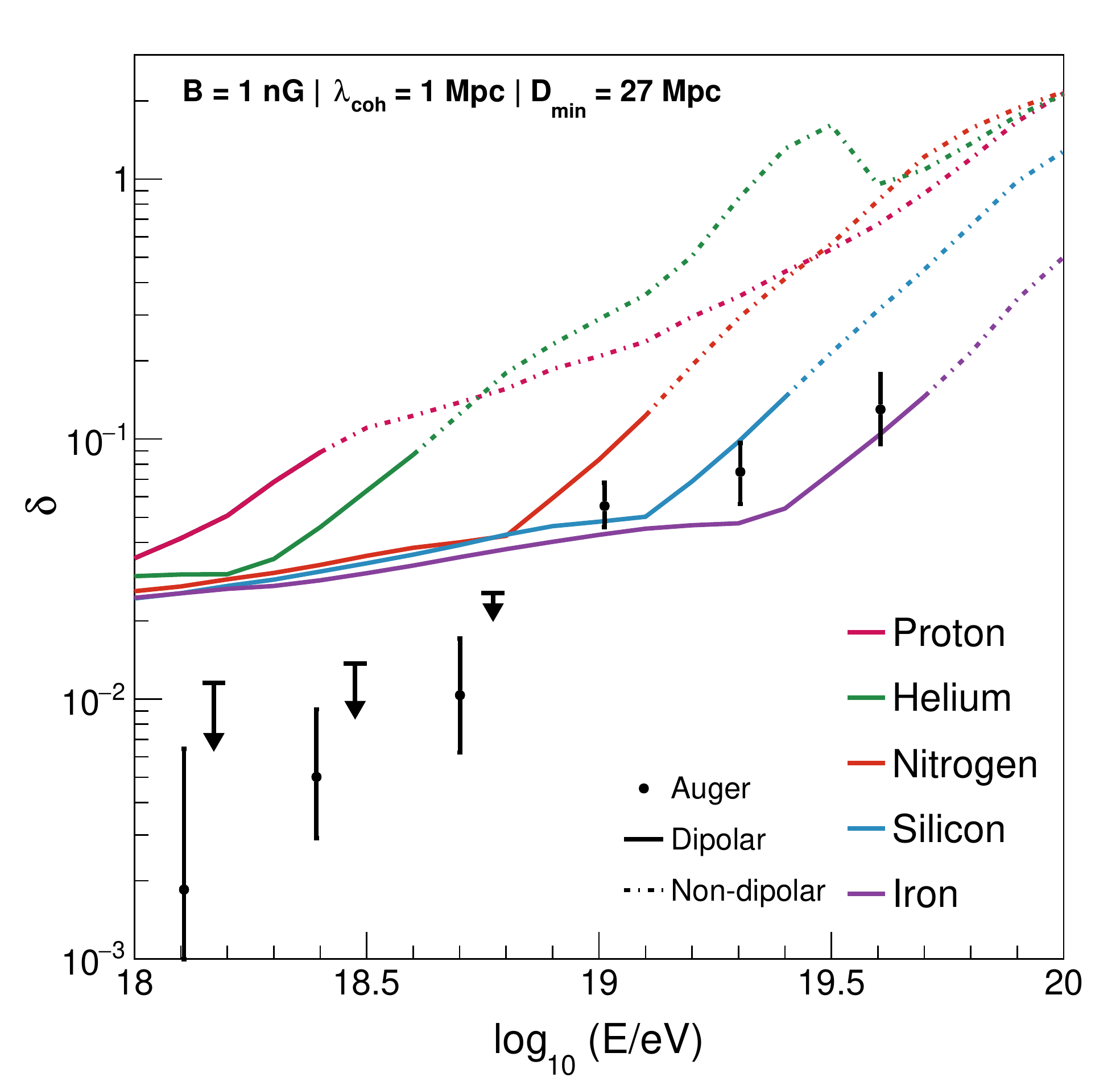}
    \caption{Evolution of the amplitude of the dipole with energy. The data points show the amplitude of the large-scale dipole measured by the Pierre Auger Observatory~\cite{AugerRightAscension}. The continuous lines show the region in which the angular distribution is dipolar, i.e., $\fnd < 1\%$, while the dashed lines show the region in which the non-dipolarity is significant.}
    \label{fig:dipole2}
\end{figure}

\begin{figure}[ht]
    \centering
    \includegraphics[width=0.49\textwidth]{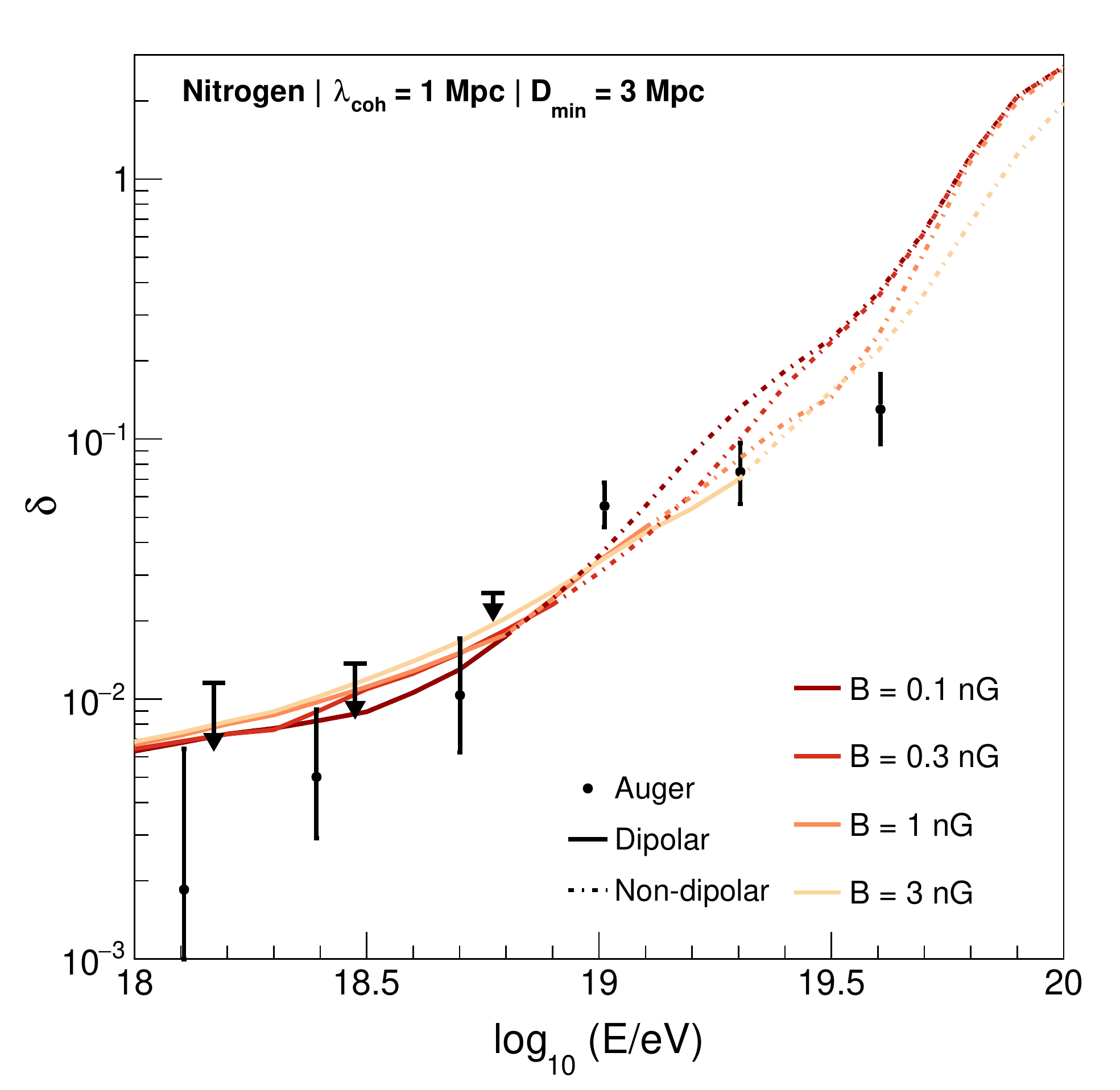}
    \caption{Same as figure~\ref{fig:dipole2}, but for a fixed distance to the nearest source and primary and different magnetic field intensities.}
    \label{fig:dipole3}
\end{figure}

\begin{figure*}[ht]
    \centering
    \includegraphics[width=0.4\textwidth]{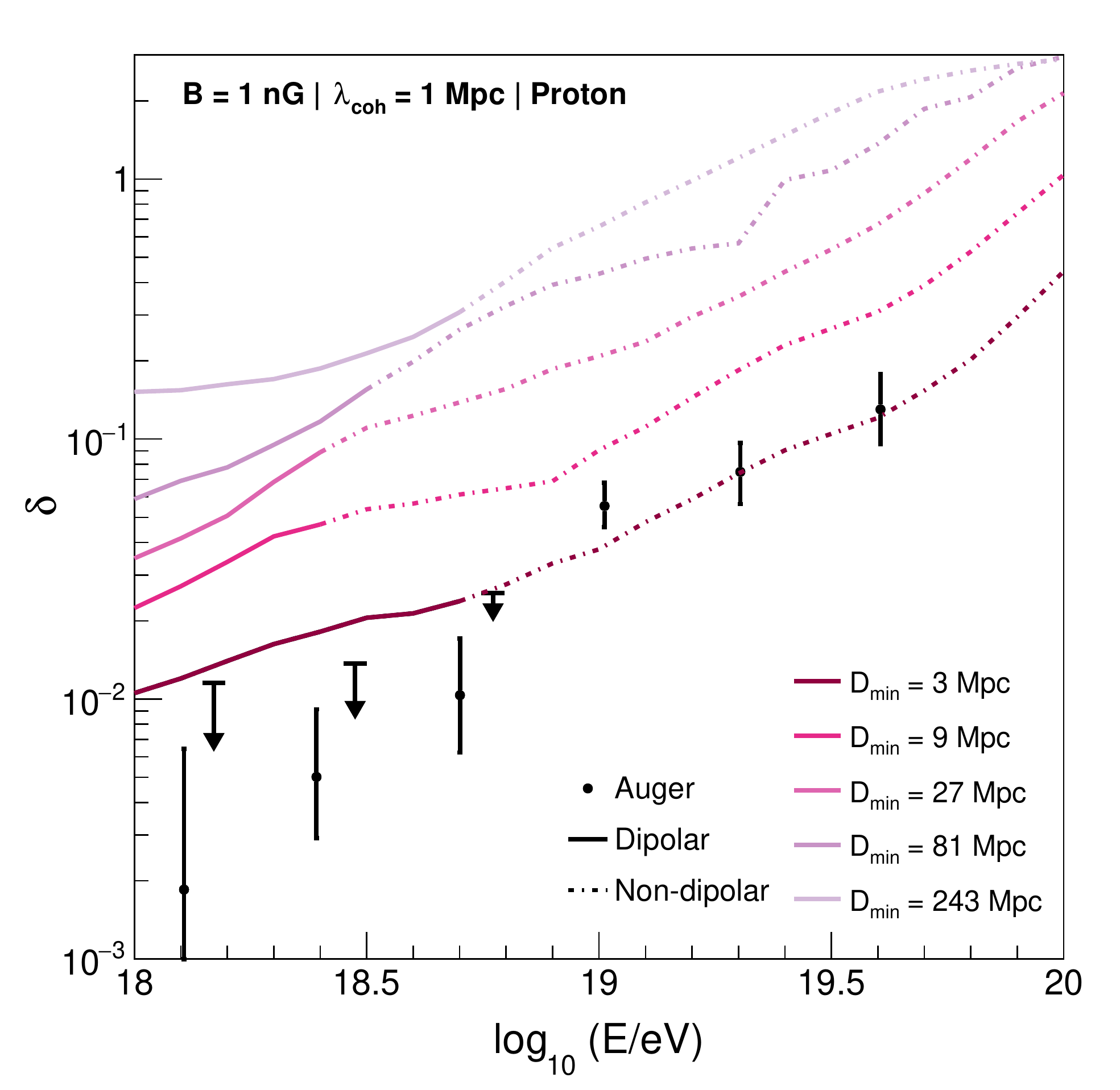}
    \includegraphics[width=0.4\textwidth]{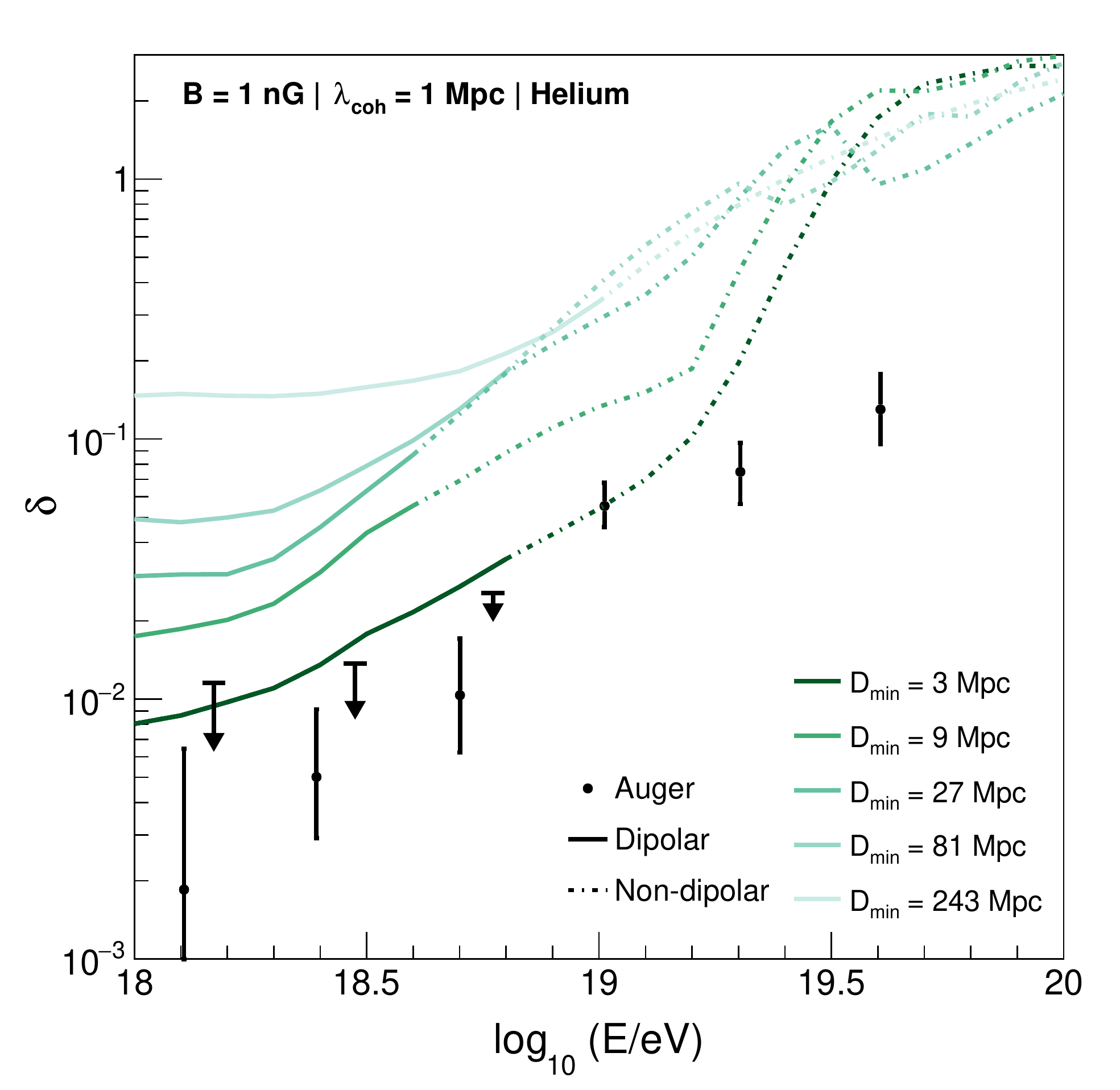}
    \includegraphics[width=0.4\textwidth]{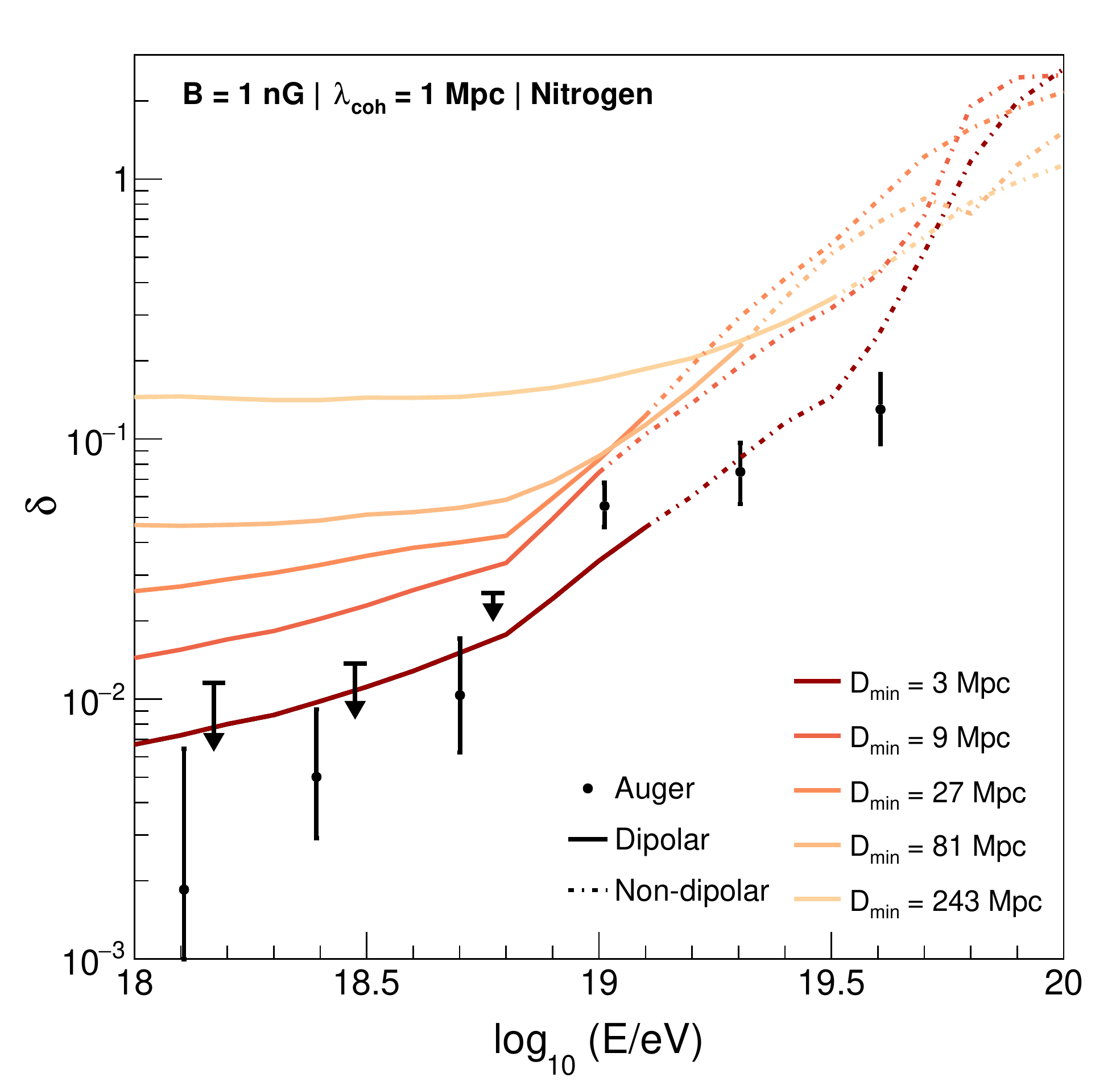}
    \includegraphics[width=0.4\textwidth]{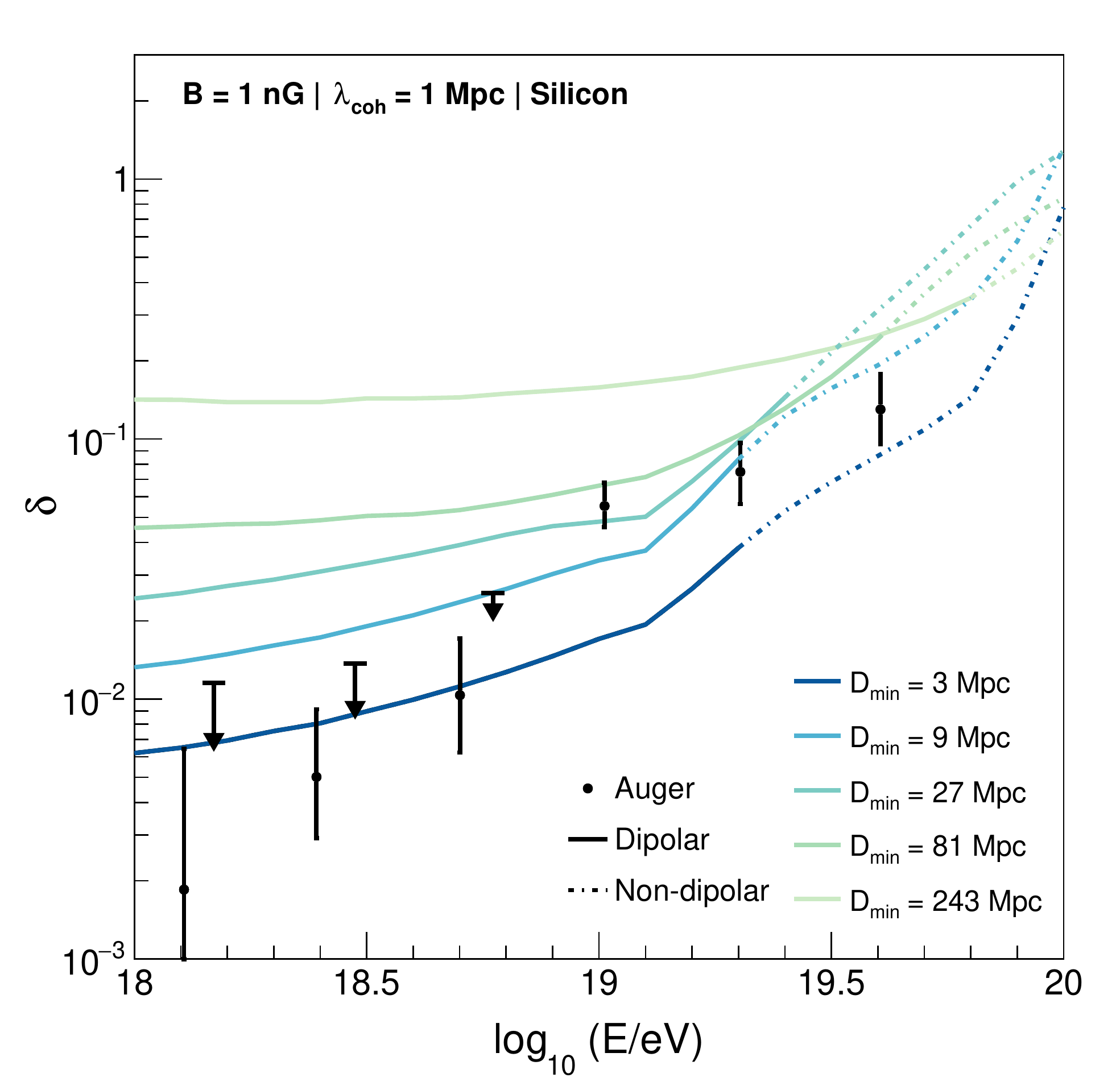}
    \includegraphics[width=0.4\textwidth]{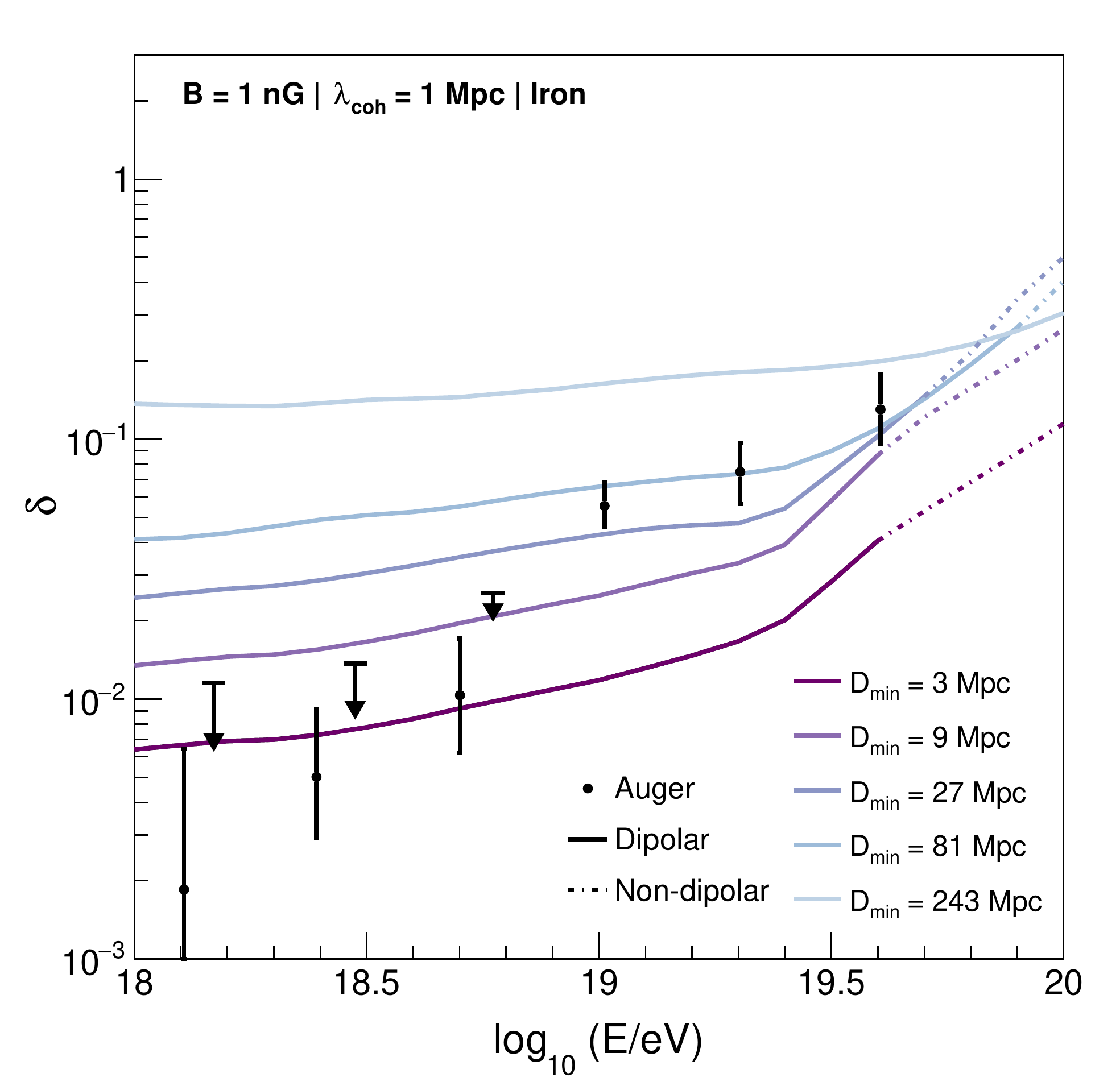}
    \caption{Same as figure~\ref{fig:dipole2}, but for a fixed magnetic field intensity and primary. Each line represents a distance to the nearest source and each panel represents a primary.}
    \label{fig:dipole1}
\end{figure*}

Figure~\ref{fig:dipole2} shows the results for different species, for the case of a fixed distance to the nearest source and scattering length. For the cases considered, up to four regimes can be seen, which directly relate to the regimes on the contribution to the spectrum of sources from a distance shell (see Ref.~\cite{LangPRD2020} for a detailed explanation of these regimes). These regimes can be divided into the non-dipolar (\textbf{A}, $\fnd > 1\%$) and three subregimes within the dipolar (\textbf{B}, $\fnd < 1\%$). From higher to lower energies: 

\begin{itemize}
\item \textbf{AI - Non-dipolar regime:} at the highest energies, the angular distribution is dominated by the closest sources, since the energy loss horizon prevents farther sources contributing to the arriving cosmic rays. These sources are in the ballistic regime and, therefore, the total angular distribution is non-dipolar, i.e., $\fnd > 1\%$;
\item \textbf{BI - Non-resonant scattering enhancement:} the dipole is dictated by the first source and diluted by the farther ones. In this regime $\lsca$ grows as $E^2$, which is reflected in the dipole;
\item \textbf{BII - Resonant scattering enhancement:} same as the previous one, but $\lsca$, and consequently the dipole grow as $E^{1/3}$;
\item \textbf{BIII - Low energy magnetic horizon:} in this regime the behavior of the dipole is dictated by the finite age of the universe, which leads to the magnetic horizon.
\end{itemize}

Figure~\ref{fig:dipole3} shows the dependence of the dipole strength (assuming a fixed distance to the nearest source, CR species, and magnetic field coherence length) for a range of extragalactic magnetic field strength values. For most cases, the dipole strength is seen to not depend on the extragalactic magnetic field strength, $B$. In this case, CR transport from sources reaches the steady-state diffusive regime value, in which the dipole grows linearly with $\lsca$, whilst the contribution to the total CR density from the closest source grows with $1/\lsca$, resulting in less dilution. These two effects subsequently compensate each other. For larger distances, however, this is not true anymore since the effects of a finite source activity time, $\tmax$, become relevant. Also, as expected, the energy at which the non-dipolar regime becomes dominant heavily depends on the magnetic field strength, $B$.

Finally, figure~\ref{fig:dipole1} shows the dipole dependence (assuming a fixed extragalactic magnetic field strength and coherence length) for different cosmic ray primary species and distances to the nearest source. The dependence on $D_{\rm min}$ follows from the fact that the dipole of the first source depends on  $1/\Dmin$, whilst the dilution factor, which depends inversely on the number of contributing sources, grows with $\Dmin^2$. Consequently, lower values of $\Dmin$ imply lower values for the dipole (ie. the dilution effect wins), at least in the energy region below the non-dipolar regime (ie. for energies in which $\fnd <1\%$).

Although it is not our aim here to make strong comparisons to the results from the Pierre Auger Observatory, a few general comments are worth noting. While for the intermediate energies ($10^{18.9}~\mathrm{eV} < E < 10^{19.4}~\mathrm{eV}$), the evolution of the dipole is consistent with an intermediate composition, the distribution at the most energetic data point ($E \sim 10^{19.6}$~eV) needs to be driven by either heavier nuclei at the sources, such as Silicon and Iron, or strong extragalactic magnetic fields ($B > 1$~nG). Furthermore, a strong evolution of the dipole strength with energy was only found for cases in which the dipole contribution from the nearest source was in the steady-state diffusion regime. Lastly, in most of the cases, the distribution becomes non-dipolar ($\fnd > 1\%$) at the highest energies, which might be consistent to the sky maps presented by the Pierre Auger Observatory in their anisotropy studies~\cite{AugerStarburst,AugerTAICRC}.

It is noteworthy that the calculations presented here consider the mean dipole strength obtained for an ensemble of randomly distributed sources. However, due to Poissonian fluctuations in the position of randomly distributed sources, a large variance of the order of the mean is expected, as estimated in appendix~\ref{app:approximation}. A more rigorous comparison with the data would, thus, be extremely dependent on the considerations about the position of the sources.

\section{Conclusions}
\label{sec:conclusions}

In this work, we have further developed a picture for the development of a large-scale dipole in the CR arrival direction distribution due to their diffusion in extragalactic magnetic fields. We have proposed a simple general approach for calculating the angular distribution, and subsequently the dipole and angular power spectrum of arriving cosmic rays emanating from an ensemble of sources.

The dependency of the dipole and angular power spectrum on the particle scattering length, source distance, and activity age the sources, for a range of transport regimes, were found to have a simple origin. For intermediate values of $\lsca/r_s$ ($r_s/\tmax \lesssim \lsca/r_s \lesssim 1$), the steady-state diffusive regime result in which the dipole behaves as $\delta \approx \lsca/r_s$, was verified. For lower values of $\lsca/r_s$, the onset of a magnetic horizon which increases the dipole was shown. For large values of $\lsca/r_s$, the transition to the ballistic regime was shown to be dominant, and a quantification of the non-dipolarity of the distribution was proposed in the form of the term $\fnd$.

The combination of the general approach for the angular distribution of a single source proposed in this work with the semi-analytical method for the propagation of UHECR proposed in our previous work~\cite{LangPRD2020} provides the necessary tools for studying the evolution with energy of the dipole from an ensemble of sources. Using this tool, a novel approach taking into account different contributions from each source due to energy losses and extragalactic magnetic fields effects as well as different primaries becomes possible.

We have covered and discussed the different regimes for the dipole strength evolution with energy, for different combinations of extragalactic magnetic field strength value, distance to the nearest source, age of activity of the sources and primary cosmic ray species.
Investigating the effect of changing the distance of the nearest source, the dipole was shown to be dictated by the closest sources, while diluted by the farther ones. From that, the distance to nearest source, $\Dmin$, was found to control the amplitude of the dipole for the energies where the arriving distribution is dipolar. Smaller values of $\Dmin$ gave rise to a larger number of contributing sources, acting overall to dilute the dipole and, consequently, reduce the overall dipole amplitude. Also, a steep energy evolution of the dipole is only found when steady-state diffusion regime is achieved. As shown in fig.~\ref{fig:dipole1}, this happens only if small values of $\Dmin$ are considered. This general result further supports the previous findings that a local UHECR source must exist \cite{Taylor11,Biermann_2012,LangPRD2020}.

The transition from a dipolar to a non-dipolar distribution and its importance to modeling the anistropy data measured by the Pierre Auger Observatory were also discussed via an angular power spectrum analysis. We have shown that, for some realistic scenarios, the distribution is already expected to be significantly non-dipolar at the highest energies. For these cases, a simple comparison of the amplitude of the calculated dipole to the dipole measured by the Pierre Auger Observatory is not consistent. A combination of more realistic models and further data on the higher poles of the distribution measured by the Pierre Auger Observatory as well as their evolution with energy (similarly to what has been done in Ref.~\cite{Aartsen:2018ppz}) can prove to be an essential key to decipher this question. A statistical comparison of the model developed here with data is outside of the scope of this work. The results here presented contribute to building up the understanding of the arrival direction of UHECRs and to providing important insights for testing realistic models.

In summary, we have improved our understanding on the dipolar behavior in the distribution of arrival directions of UHECR and provided a relevant set of tools for pursuing the answer for the century-long question about their origin.

\section*{Acknowledgements}
RGL and VdS acknowledge FAPESP support No.
2015/15897-1, No. 2016/24943-0 and No. 2019/01653-4. RGL and VdS acknowledge the National Laboratory for Scientific Computing (LNCC/MCTI, Brazil) for providing HPC resources of the SDumont supercomputer, which have contributed to the research results reported within this paper. (\url{http://sdumont.lncc.br}). RGL thanks DESY Zeuthen for all the help and infra-structure provided while visiting the institution. VdS thanks CNPq.

\bibliography{references.bib}

\begin{thebibliography}{32}%
\makeatletter
\providecommand \@ifxundefined [1]{%
 \@ifx{#1\undefined}
}%
\providecommand \@ifnum [1]{%
 \ifnum #1\expandafter \@firstoftwo
 \else \expandafter \@secondoftwo
 \fi
}%
\providecommand \@ifx [1]{%
 \ifx #1\expandafter \@firstoftwo
 \else \expandafter \@secondoftwo
 \fi
}%
\providecommand \natexlab [1]{#1}%
\providecommand \enquote  [1]{``#1''}%
\providecommand \bibnamefont  [1]{#1}%
\providecommand \bibfnamefont [1]{#1}%
\providecommand \citenamefont [1]{#1}%
\providecommand \href@noop [0]{\@secondoftwo}%
\providecommand \href [0]{\begingroup \@sanitize@url \@href}%
\providecommand \@href[1]{\@@startlink{#1}\@@href}%
\providecommand \@@href[1]{\endgroup#1\@@endlink}%
\providecommand \@sanitize@url [0]{\catcode `\\12\catcode `\$12\catcode
  `\&12\catcode `\#12\catcode `\^12\catcode `\_12\catcode `\%12\relax}%
\providecommand \@@startlink[1]{}%
\providecommand \@@endlink[0]{}%
\providecommand \url  [0]{\begingroup\@sanitize@url \@url }%
\providecommand \@url [1]{\endgroup\@href {#1}{\urlprefix }}%
\providecommand \urlprefix  [0]{URL }%
\providecommand \Eprint [0]{\href }%
\providecommand \doibase [0]{http://dx.doi.org/}%
\providecommand \selectlanguage [0]{\@gobble}%
\providecommand \bibinfo  [0]{\@secondoftwo}%
\providecommand \bibfield  [0]{\@secondoftwo}%
\providecommand \translation [1]{[#1]}%
\providecommand \BibitemOpen [0]{}%
\providecommand \bibitemStop [0]{}%
\providecommand \bibitemNoStop [0]{.\EOS\space}%
\providecommand \EOS [0]{\spacefactor3000\relax}%
\providecommand \BibitemShut  [1]{\csname bibitem#1\endcsname}%
\let\auto@bib@innerbib\@empty
%</preamble>
\bibitem [{\citenamefont {Alves~Batista}\ \emph {et~al.}(2019)\citenamefont
  {Alves~Batista} \emph {et~al.}}]{AlvesBatista:2019tlv}%
  \BibitemOpen
  \bibfield  {author} {\bibinfo {author} {\bibfnamefont {R.}~\bibnamefont
  {Alves~Batista}} \emph {et~al.},\ }\href {\doibase 10.3389/fspas.2019.00023}
  {\bibfield  {journal} {\bibinfo  {journal} {Front. Astron. Space Sci.}\
  }\textbf {\bibinfo {volume} {6}},\ \bibinfo {pages} {23} (\bibinfo {year}
  {2019})},\ \Eprint {http://arxiv.org/abs/1903.06714} {arXiv:1903.06714
  [astro-ph.HE]} \BibitemShut {NoStop}%
%%CITATION = ARXIV:1903.06714;%%
\bibitem [{\citenamefont {Aab}\ \emph {et~al.}(2015)\citenamefont {Aab} \emph
  {et~al.}}]{PierreAuger}%
  \BibitemOpen
  \bibfield  {author} {\bibinfo {author} {\bibfnamefont {A.}~\bibnamefont
  {Aab}} \emph {et~al.} (\bibinfo {collaboration} {Pierre Auger}),\ }\href
  {\doibase 10.1016/j.nima.2015.06.058} {\bibfield  {journal} {\bibinfo
  {journal} {Nucl. Instrum. Meth. A}\ }\textbf {\bibinfo {volume} {798}},\
  \bibinfo {pages} {172} (\bibinfo {year} {2015})},\ \Eprint
  {http://arxiv.org/abs/1502.01323} {arXiv:1502.01323 [astro-ph.IM]}
  \BibitemShut {NoStop}%
\bibitem [{\citenamefont {Aab}\ \emph {et~al.}(2017)\citenamefont {Aab} \emph
  {et~al.}}]{AugerDipole}%
  \BibitemOpen
  \bibfield  {author} {\bibinfo {author} {\bibfnamefont {A.}~\bibnamefont
  {Aab}} \emph {et~al.} (\bibinfo {collaboration} {Pierre Auger}),\ }\href
  {\doibase 10.1126/science.aan4338} {\bibfield  {journal} {\bibinfo  {journal}
  {Science}\ }\textbf {\bibinfo {volume} {357}},\ \bibinfo {pages} {1266}
  (\bibinfo {year} {2017})},\ \Eprint {http://arxiv.org/abs/1709.07321}
  {arXiv:1709.07321 [astro-ph.HE]} \BibitemShut {NoStop}%
\bibitem [{\citenamefont {Aab}\ \emph {et~al.}(2020{\natexlab{a}})\citenamefont
  {Aab} \emph {et~al.}}]{AugerRightAscension}%
  \BibitemOpen
  \bibfield  {author} {\bibinfo {author} {\bibfnamefont {A.}~\bibnamefont
  {Aab}} \emph {et~al.} (\bibinfo {collaboration} {Pierre Auger}),\ }\href
  {\doibase 10.3847/1538-4357/ab7236} {\bibfield  {journal} {\bibinfo
  {journal} {Astrophys. J.}\ }\textbf {\bibinfo {volume} {891}},\ \bibinfo
  {pages} {142} (\bibinfo {year} {2020}{\natexlab{a}})},\ \Eprint
  {http://arxiv.org/abs/2002.06172} {arXiv:2002.06172 [astro-ph.HE]}
  \BibitemShut {NoStop}%
\bibitem [{\citenamefont {Abraham}\ \emph {et~al.}(2010)\citenamefont {Abraham}
  \emph {et~al.}}]{AugerSpectrum}%
  \BibitemOpen
  \bibfield  {author} {\bibinfo {author} {\bibfnamefont {J.}~\bibnamefont
  {Abraham}} \emph {et~al.} (\bibinfo {collaboration} {Pierre Auger}),\ }\href
  {\doibase 10.1016/j.physletb.2010.02.013} {\bibfield  {journal} {\bibinfo
  {journal} {Phys. Lett. B}\ }\textbf {\bibinfo {volume} {685}},\ \bibinfo
  {pages} {239} (\bibinfo {year} {2010})},\ \Eprint
  {http://arxiv.org/abs/1002.1975} {arXiv:1002.1975 [astro-ph.HE]} \BibitemShut
  {NoStop}%
\bibitem [{\citenamefont {Aab}\ \emph {et~al.}(2020{\natexlab{b}})\citenamefont
  {Aab} \emph {et~al.}}]{AugerSpectrum1}%
  \BibitemOpen
  \bibfield  {author} {\bibinfo {author} {\bibfnamefont {A.}~\bibnamefont
  {Aab}} \emph {et~al.} (\bibinfo {collaboration} {Pierre Auger}),\ }\href
  {\doibase 10.1103/PhysRevLett.125.121106} {\bibfield  {journal} {\bibinfo
  {journal} {Phys. Rev. Lett.}\ }\textbf {\bibinfo {volume} {125}},\ \bibinfo
  {pages} {121106} (\bibinfo {year} {2020}{\natexlab{b}})},\ \Eprint
  {http://arxiv.org/abs/2008.06488} {arXiv:2008.06488 [astro-ph.HE]}
  \BibitemShut {NoStop}%
\bibitem [{\citenamefont {Aab}\ \emph {et~al.}(2020{\natexlab{c}})\citenamefont
  {Aab} \emph {et~al.}}]{AugerSpectrum2}%
  \BibitemOpen
  \bibfield  {author} {\bibinfo {author} {\bibfnamefont {A.}~\bibnamefont
  {Aab}} \emph {et~al.} (\bibinfo {collaboration} {Pierre Auger}),\ }\href
  {\doibase 10.1103/PhysRevD.102.062005} {\bibfield  {journal} {\bibinfo
  {journal} {Phys. Rev. D}\ }\textbf {\bibinfo {volume} {102}},\ \bibinfo
  {pages} {062005} (\bibinfo {year} {2020}{\natexlab{c}})},\ \Eprint
  {http://arxiv.org/abs/2008.06486} {arXiv:2008.06486 [astro-ph.HE]}
  \BibitemShut {NoStop}%
\bibitem [{\citenamefont {Lang}\ \emph {et~al.}(2020)\citenamefont {Lang},
  \citenamefont {Taylor}, \citenamefont {Ahlers},\ and\ \citenamefont
  {de~Souza}}]{LangPRD2020}%
  \BibitemOpen
  \bibfield  {author} {\bibinfo {author} {\bibfnamefont {R.~G.}\ \bibnamefont
  {Lang}}, \bibinfo {author} {\bibfnamefont {A.~M.}\ \bibnamefont {Taylor}},
  \bibinfo {author} {\bibfnamefont {M.}~\bibnamefont {Ahlers}}, \ and\ \bibinfo
  {author} {\bibfnamefont {V.}~\bibnamefont {de~Souza}},\ }\href {\doibase
  10.1103/PhysRevD.102.063012} {\bibfield  {journal} {\bibinfo  {journal}
  {Phys. Rev. D}\ }\textbf {\bibinfo {volume} {102}},\ \bibinfo {pages}
  {063012} (\bibinfo {year} {2020})}\BibitemShut {NoStop}%
\bibitem [{\citenamefont {{Schlickeiser}}(1989)}]{1989ApJ...336..243S}%
  \BibitemOpen
  \bibfield  {author} {\bibinfo {author} {\bibfnamefont {R.}~\bibnamefont
  {{Schlickeiser}}},\ }\href {\doibase 10.1086/167009} {\bibfield  {journal}
  {\bibinfo  {journal} {\apj}\ }\textbf {\bibinfo {volume} {336}},\ \bibinfo
  {pages} {243} (\bibinfo {year} {1989})}\BibitemShut {NoStop}%
\bibitem [{\citenamefont {O'Sullivan}\ \emph {et~al.}(2009)\citenamefont
  {O'Sullivan}, \citenamefont {Reville},\ and\ \citenamefont
  {Taylor}}]{OSullivan:2009rvg}%
  \BibitemOpen
  \bibfield  {author} {\bibinfo {author} {\bibfnamefont {S.}~\bibnamefont
  {O'Sullivan}}, \bibinfo {author} {\bibfnamefont {B.}~\bibnamefont {Reville}},
  \ and\ \bibinfo {author} {\bibfnamefont {A.}~\bibnamefont {Taylor}},\ }\href
  {\doibase 10.1111/j.1365-2966.2009.15442.x} {\bibfield  {journal} {\bibinfo
  {journal} {Mon. Not. Roy. Astron. Soc.}\ }\textbf {\bibinfo {volume} {400}},\
  \bibinfo {pages} {248} (\bibinfo {year} {2009})},\ \Eprint
  {http://arxiv.org/abs/0903.1259} {arXiv:0903.1259 [astro-ph.HE]} \BibitemShut
  {NoStop}%
\bibitem [{\citenamefont {Harari}\ \emph {et~al.}(2014)\citenamefont {Harari},
  \citenamefont {Mollerach},\ and\ \citenamefont {Roulet}}]{Harari:2013pea}%
  \BibitemOpen
  \bibfield  {author} {\bibinfo {author} {\bibfnamefont {D.}~\bibnamefont
  {Harari}}, \bibinfo {author} {\bibfnamefont {S.}~\bibnamefont {Mollerach}}, \
  and\ \bibinfo {author} {\bibfnamefont {E.}~\bibnamefont {Roulet}},\ }\href
  {\doibase 10.1103/PhysRevD.89.123001} {\bibfield  {journal} {\bibinfo
  {journal} {Phys. Rev. D}\ }\textbf {\bibinfo {volume} {89}},\ \bibinfo
  {pages} {123001} (\bibinfo {year} {2014})},\ \Eprint
  {http://arxiv.org/abs/1312.1366} {arXiv:1312.1366 [astro-ph.HE]} \BibitemShut
  {NoStop}%
\bibitem [{\citenamefont {Ahlers}\ and\ \citenamefont
  {Mertsch}(2017)}]{Ahlers:2016rox}%
  \BibitemOpen
  \bibfield  {author} {\bibinfo {author} {\bibfnamefont {M.}~\bibnamefont
  {Ahlers}}\ and\ \bibinfo {author} {\bibfnamefont {P.}~\bibnamefont
  {Mertsch}},\ }\href {\doibase 10.1016/j.ppnp.2017.01.004} {\bibfield
  {journal} {\bibinfo  {journal} {Prog. Part. Nucl. Phys.}\ }\textbf {\bibinfo
  {volume} {94}},\ \bibinfo {pages} {184} (\bibinfo {year} {2017})},\ \Eprint
  {http://arxiv.org/abs/1612.01873} {arXiv:1612.01873 [astro-ph.HE]}
  \BibitemShut {NoStop}%
\bibitem [{\citenamefont {Wittkowski}\ and\ \citenamefont
  {Kampert}(2018)}]{Wittkowski:2017nfd}%
  \BibitemOpen
  \bibfield  {author} {\bibinfo {author} {\bibfnamefont {D.}~\bibnamefont
  {Wittkowski}}\ and\ \bibinfo {author} {\bibfnamefont {K.-H.}\ \bibnamefont
  {Kampert}},\ }\href {\doibase 10.3847/2041-8213/aaa2f9} {\bibfield  {journal}
  {\bibinfo  {journal} {Astrophys. J. Lett.}\ }\textbf {\bibinfo {volume}
  {854}},\ \bibinfo {pages} {L3} (\bibinfo {year} {2018})},\ \Eprint
  {http://arxiv.org/abs/1710.05617} {arXiv:1710.05617 [astro-ph.HE]}
  \BibitemShut {NoStop}%
\bibitem [{\citenamefont {Abeysekara}\ \emph {et~al.}(2019)\citenamefont
  {Abeysekara} \emph {et~al.}}]{Aartsen:2018ppz}%
  \BibitemOpen
  \bibfield  {author} {\bibinfo {author} {\bibfnamefont {A.}~\bibnamefont
  {Abeysekara}} \emph {et~al.} (\bibinfo {collaboration} {HAWC, IceCube}),\
  }\href {\doibase 10.3847/1538-4357/aaf5cc} {\bibfield  {journal} {\bibinfo
  {journal} {Astrophys. J.}\ }\textbf {\bibinfo {volume} {871}},\ \bibinfo
  {pages} {96} (\bibinfo {year} {2019})},\ \Eprint
  {http://arxiv.org/abs/1812.05682} {arXiv:1812.05682 [astro-ph.HE]}
  \BibitemShut {NoStop}%
\bibitem [{\citenamefont {Harari}\ \emph {et~al.}(2015)\citenamefont {Harari},
  \citenamefont {Mollerach},\ and\ \citenamefont {Roulet}}]{Harari:2015hba}%
  \BibitemOpen
  \bibfield  {author} {\bibinfo {author} {\bibfnamefont {D.}~\bibnamefont
  {Harari}}, \bibinfo {author} {\bibfnamefont {S.}~\bibnamefont {Mollerach}}, \
  and\ \bibinfo {author} {\bibfnamefont {E.}~\bibnamefont {Roulet}},\ }\href
  {\doibase 10.1103/PhysRevD.92.063014} {\bibfield  {journal} {\bibinfo
  {journal} {Phys. Rev. D}\ }\textbf {\bibinfo {volume} {92}},\ \bibinfo
  {pages} {063014} (\bibinfo {year} {2015})},\ \Eprint
  {http://arxiv.org/abs/1507.06585} {arXiv:1507.06585 [astro-ph.HE]}
  \BibitemShut {NoStop}%
\bibitem [{\citenamefont {Globus}\ and\ \citenamefont
  {Piran}(2017)}]{Globus:2017fym}%
  \BibitemOpen
  \bibfield  {author} {\bibinfo {author} {\bibfnamefont {N.}~\bibnamefont
  {Globus}}\ and\ \bibinfo {author} {\bibfnamefont {T.}~\bibnamefont {Piran}},\
  }\href {\doibase 10.3847/2041-8213/aa991b} {\bibfield  {journal} {\bibinfo
  {journal} {Astrophys. J. Lett.}\ }\textbf {\bibinfo {volume} {850}},\
  \bibinfo {pages} {L25} (\bibinfo {year} {2017})},\ \Eprint
  {http://arxiv.org/abs/1709.10110} {arXiv:1709.10110 [astro-ph.HE]}
  \BibitemShut {NoStop}%
\bibitem [{\citenamefont {Dundovi\'c}\ and\ \citenamefont
  {Sigl}(2019)}]{Dundovic:2017vsz}%
  \BibitemOpen
  \bibfield  {author} {\bibinfo {author} {\bibfnamefont {A.}~\bibnamefont
  {Dundovi\'c}}\ and\ \bibinfo {author} {\bibfnamefont {G.}~\bibnamefont
  {Sigl}},\ }\href {\doibase 10.1088/1475-7516/2019/01/018} {\bibfield
  {journal} {\bibinfo  {journal} {JCAP}\ }\textbf {\bibinfo {volume} {01}},\
  \bibinfo {pages} {018} (\bibinfo {year} {2019})},\ \Eprint
  {http://arxiv.org/abs/1710.05517} {arXiv:1710.05517 [astro-ph.HE]}
  \BibitemShut {NoStop}%
\bibitem [{\citenamefont {Mollerach}\ and\ \citenamefont
  {Roulet}(2020)}]{Mollerach:2020mhr}%
  \BibitemOpen
  \bibfield  {author} {\bibinfo {author} {\bibfnamefont {S.}~\bibnamefont
  {Mollerach}}\ and\ \bibinfo {author} {\bibfnamefont {E.}~\bibnamefont
  {Roulet}},\ }\href {\doibase 10.1103/PhysRevD.101.103024} {\bibfield
  {journal} {\bibinfo  {journal} {Phys. Rev. D}\ }\textbf {\bibinfo {volume}
  {101}},\ \bibinfo {pages} {103024} (\bibinfo {year} {2020})},\ \Eprint
  {http://arxiv.org/abs/2004.04253} {arXiv:2004.04253 [astro-ph.HE]}
  \BibitemShut {NoStop}%
\bibitem [{\citenamefont {Hooper}\ \emph {et~al.}(2007)\citenamefont {Hooper},
  \citenamefont {Sarkar},\ and\ \citenamefont {Taylor}}]{Hooper:2006tn}%
  \BibitemOpen
  \bibfield  {author} {\bibinfo {author} {\bibfnamefont {D.}~\bibnamefont
  {Hooper}}, \bibinfo {author} {\bibfnamefont {S.}~\bibnamefont {Sarkar}}, \
  and\ \bibinfo {author} {\bibfnamefont {A.~M.}\ \bibnamefont {Taylor}},\
  }\href {\doibase 10.1016/j.astropartphys.2006.10.008} {\bibfield  {journal}
  {\bibinfo  {journal} {Astropart. Phys.}\ }\textbf {\bibinfo {volume} {27}},\
  \bibinfo {pages} {199} (\bibinfo {year} {2007})},\ \Eprint
  {http://arxiv.org/abs/astro-ph/0608085} {arXiv:astro-ph/0608085 [astro-ph]}
  \BibitemShut {NoStop}%
%%CITATION = ASTRO-PH/0608085;%%
\bibitem [{\citenamefont {Allard}(2012)}]{Allard:2011aa}%
  \BibitemOpen
  \bibfield  {author} {\bibinfo {author} {\bibfnamefont {D.}~\bibnamefont
  {Allard}},\ }\href {\doibase 10.1016/j.astropartphys.2011.10.011} {\bibfield
  {journal} {\bibinfo  {journal} {Astropart. Phys.}\ }\textbf {\bibinfo
  {volume} {39-40}},\ \bibinfo {pages} {33} (\bibinfo {year} {2012})},\ \Eprint
  {http://arxiv.org/abs/1111.3290} {arXiv:1111.3290 [astro-ph.HE]} \BibitemShut
  {NoStop}%
%%CITATION = ARXIV:1111.3290;%%
\bibitem [{\citenamefont {Lemoine}(2005)}]{Lemoine_2005}%
  \BibitemOpen
  \bibfield  {author} {\bibinfo {author} {\bibfnamefont {M.}~\bibnamefont
  {Lemoine}},\ }\href {\doibase 10.1103/PhysRevD.71.083007} {\bibfield
  {journal} {\bibinfo  {journal} {Phys. Rev. D}\ }\textbf {\bibinfo {volume}
  {71}},\ \bibinfo {pages} {083007} (\bibinfo {year} {2005})},\ \Eprint
  {http://arxiv.org/abs/astro-ph/0411173} {arXiv:astro-ph/0411173} \BibitemShut
  {NoStop}%
\bibitem [{\citenamefont {Aloisio}\ and\ \citenamefont
  {Berezinsky}(2005)}]{Aloisio_2005}%
  \BibitemOpen
  \bibfield  {author} {\bibinfo {author} {\bibfnamefont {R.}~\bibnamefont
  {Aloisio}}\ and\ \bibinfo {author} {\bibfnamefont {V.}~\bibnamefont
  {Berezinsky}},\ }\href {\doibase 10.1086/429615} {\bibfield  {journal}
  {\bibinfo  {journal} {Astrophys. J.}\ }\textbf {\bibinfo {volume} {625}},\
  \bibinfo {pages} {249} (\bibinfo {year} {2005})},\ \Eprint
  {http://arxiv.org/abs/astro-ph/0412578} {arXiv:astro-ph/0412578} \BibitemShut
  {NoStop}%
\bibitem [{\citenamefont {Globus}\ \emph {et~al.}(2008)\citenamefont {Globus},
  \citenamefont {Allard},\ and\ \citenamefont {Parizot}}]{Globus_2007}%
  \BibitemOpen
  \bibfield  {author} {\bibinfo {author} {\bibfnamefont {N.}~\bibnamefont
  {Globus}}, \bibinfo {author} {\bibfnamefont {D.}~\bibnamefont {Allard}}, \
  and\ \bibinfo {author} {\bibfnamefont {E.}~\bibnamefont {Parizot}},\ }\href
  {\doibase 10.1051/0004-6361:20078653} {\bibfield  {journal} {\bibinfo
  {journal} {Astron. Astrophys.}\ }\textbf {\bibinfo {volume} {479}},\ \bibinfo
  {pages} {97} (\bibinfo {year} {2008})},\ \Eprint
  {http://arxiv.org/abs/0709.1541} {arXiv:0709.1541 [astro-ph]} \BibitemShut
  {NoStop}%
\bibitem [{\citenamefont {Taylor}\ \emph {et~al.}(2011)\citenamefont {Taylor},
  \citenamefont {Ahlers},\ and\ \citenamefont {Aharonian}}]{Taylor11}%
  \BibitemOpen
  \bibfield  {author} {\bibinfo {author} {\bibfnamefont {A.~M.}\ \bibnamefont
  {Taylor}}, \bibinfo {author} {\bibfnamefont {M.}~\bibnamefont {Ahlers}}, \
  and\ \bibinfo {author} {\bibfnamefont {F.~A.}\ \bibnamefont {Aharonian}},\
  }\href {\doibase 10.1103/PhysRevD.84.105007} {\bibfield  {journal} {\bibinfo
  {journal} {Phys. Rev. D}\ }\textbf {\bibinfo {volume} {84}},\ \bibinfo
  {pages} {105007} (\bibinfo {year} {2011})},\ \Eprint
  {http://arxiv.org/abs/1107.2055} {arXiv:1107.2055 [astro-ph.HE]} \BibitemShut
  {NoStop}%
\bibitem [{\citenamefont {Mollerach}\ and\ \citenamefont
  {Roulet}(2013)}]{Mollerach_2013}%
  \BibitemOpen
  \bibfield  {author} {\bibinfo {author} {\bibfnamefont {S.}~\bibnamefont
  {Mollerach}}\ and\ \bibinfo {author} {\bibfnamefont {E.}~\bibnamefont
  {Roulet}},\ }\href {\doibase 10.1088/1475-7516/2013/10/013} {\bibfield
  {journal} {\bibinfo  {journal} {JCAP}\ }\textbf {\bibinfo {volume} {10}},\
  \bibinfo {pages} {013} (\bibinfo {year} {2013})},\ \Eprint
  {http://arxiv.org/abs/1305.6519} {arXiv:1305.6519 [astro-ph.HE]} \BibitemShut
  {NoStop}%
\bibitem [{\citenamefont {Kronberg}\ and\ \citenamefont
  {Simard-Normandin}(1976)}]{Kronberg_1976}%
  \BibitemOpen
  \bibfield  {author} {\bibinfo {author} {\bibfnamefont {P.~P.}\ \bibnamefont
  {Kronberg}}\ and\ \bibinfo {author} {\bibfnamefont {M.}~\bibnamefont
  {Simard-Normandin}},\ }\href {\doibase 10.1038/263653a0} {\bibfield
  {journal} {\bibinfo  {journal} {Nature}\ }\textbf {\bibinfo {volume} {263}},\
  \bibinfo {pages} {653} (\bibinfo {year} {1976})}\BibitemShut {NoStop}%
\bibitem [{\citenamefont {Kronberg}(1994)}]{Kronberg:1993vk}%
  \BibitemOpen
  \bibfield  {author} {\bibinfo {author} {\bibfnamefont {P.~P.}\ \bibnamefont
  {Kronberg}},\ }\href {\doibase 10.1088/0034-4885/57/4/001} {\bibfield
  {journal} {\bibinfo  {journal} {Rept. Prog. Phys.}\ }\textbf {\bibinfo
  {volume} {57}},\ \bibinfo {pages} {325} (\bibinfo {year} {1994})}\BibitemShut
  {NoStop}%
\bibitem [{\citenamefont {Blasi}\ \emph {et~al.}(1999)\citenamefont {Blasi},
  \citenamefont {Burles},\ and\ \citenamefont {Olinto}}]{Blasi_1999}%
  \BibitemOpen
  \bibfield  {author} {\bibinfo {author} {\bibfnamefont {P.}~\bibnamefont
  {Blasi}}, \bibinfo {author} {\bibfnamefont {S.}~\bibnamefont {Burles}}, \
  and\ \bibinfo {author} {\bibfnamefont {A.~V.}\ \bibnamefont {Olinto}},\
  }\href {\doibase 10.1086/311958} {\bibfield  {journal} {\bibinfo  {journal}
  {Astrophys. J. Lett.}\ }\textbf {\bibinfo {volume} {514}},\ \bibinfo {pages}
  {L79} (\bibinfo {year} {1999})},\ \Eprint
  {http://arxiv.org/abs/astro-ph/9812487} {arXiv:astro-ph/9812487} \BibitemShut
  {NoStop}%
\bibitem [{\citenamefont {Schleicher}\ and\ \citenamefont
  {Miniati}(2011)}]{Schleicher:2011jj}%
  \BibitemOpen
  \bibfield  {author} {\bibinfo {author} {\bibfnamefont {D.~R.}\ \bibnamefont
  {Schleicher}}\ and\ \bibinfo {author} {\bibfnamefont {F.}~\bibnamefont
  {Miniati}},\ }\href {\doibase 10.1111/j.1745-3933.2011.01162.x} {\bibfield
  {journal} {\bibinfo  {journal} {Mon. Not. Roy. Astron. Soc.}\ }\textbf
  {\bibinfo {volume} {418}},\ \bibinfo {pages} {143} (\bibinfo {year}
  {2011})},\ \Eprint {http://arxiv.org/abs/1108.1874} {arXiv:1108.1874
  [astro-ph.CO]} \BibitemShut {NoStop}%
\bibitem [{\citenamefont {Aab}\ \emph {et~al.}(2018)\citenamefont {Aab} \emph
  {et~al.}}]{AugerStarburst}%
  \BibitemOpen
  \bibfield  {author} {\bibinfo {author} {\bibfnamefont {A.}~\bibnamefont
  {Aab}} \emph {et~al.} (\bibinfo {collaboration} {Pierre Auger}),\ }\href
  {\doibase 10.3847/2041-8213/aaa66d} {\bibfield  {journal} {\bibinfo
  {journal} {Astrophys. J. Lett.}\ }\textbf {\bibinfo {volume} {853}},\
  \bibinfo {pages} {L29} (\bibinfo {year} {2018})},\ \Eprint
  {http://arxiv.org/abs/1801.06160} {arXiv:1801.06160 [astro-ph.HE]}
  \BibitemShut {NoStop}%
\bibitem [{\citenamefont {di~Matteo}\ \emph {et~al.}(2020)\citenamefont
  {di~Matteo} \emph {et~al.}}]{AugerTAICRC}%
  \BibitemOpen
  \bibfield  {author} {\bibinfo {author} {\bibfnamefont {A.}~\bibnamefont
  {di~Matteo}} \emph {et~al.} (\bibinfo {collaboration} {Pierre Auger,
  Telescope Array}),\ }\href {\doibase 10.22323/1.358.0439} {\bibfield
  {journal} {\bibinfo  {journal} {PoS}\ }\textbf {\bibinfo {volume}
  {ICRC2019}},\ \bibinfo {pages} {439} (\bibinfo {year} {2020})},\ \Eprint
  {http://arxiv.org/abs/2001.01864} {arXiv:2001.01864 [astro-ph.HE]}
  \BibitemShut {NoStop}%
\bibitem [{\citenamefont {Biermann}\ and\ \citenamefont
  {de~Souza}(2012)}]{Biermann_2012}%
  \BibitemOpen
  \bibfield  {author} {\bibinfo {author} {\bibfnamefont {P.~L.}\ \bibnamefont
  {Biermann}}\ and\ \bibinfo {author} {\bibfnamefont {V.}~\bibnamefont
  {de~Souza}},\ }\href {\doibase 10.1088/0004-637X/746/1/72} {\bibfield
  {journal} {\bibinfo  {journal} {Astrophys. J.}\ }\textbf {\bibinfo {volume}
  {746}},\ \bibinfo {pages} {72} (\bibinfo {year} {2012})},\ \Eprint
  {http://arxiv.org/abs/1106.0625} {arXiv:1106.0625 [astro-ph.HE]} \BibitemShut
  {NoStop}%
\end{thebibliography}%

\begin{appendix}

\section{Dependency on age of activity of the source and angular resolution}
\label{app:singlesource}

As discussed in section~\ref{sec:singlesource}, the evolution of the amplitude of the dipole with $\lsca/r_s$ departures from the steady-state diffusive regimes for both low and high values of $\lsca/r_s$.

The horizon at lower values of $\lsca/r_s$ is related to the finite age of activity of the sources. Particles that went past Earth and must subsequently diffuse back might, in order to arrive, have a longer path than the age of the universe, in contrast with head-on particles, which have a shorter path, increasing, thus, the dipole. This regime is significant for $\lsca/r_s < r_s/c\tmax$. 

\begin{figure*}[ht]
    \centering
    \includegraphics[width=0.9\textwidth]{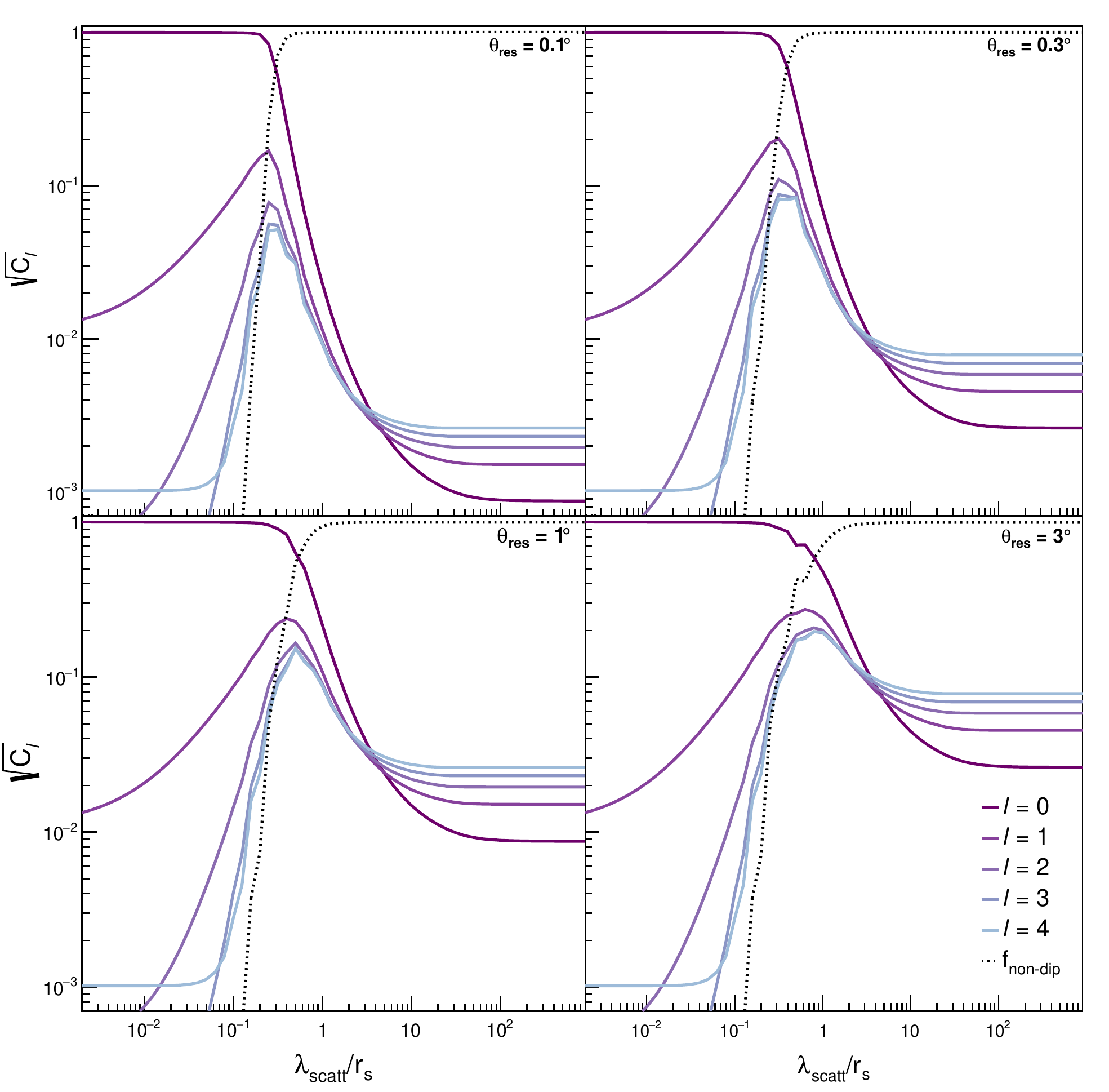}
    \caption{Evolution of the angular power spectrum with $\lambda/r_s$. The panels are for $\theta_{\mathrm{res}} = \{0.1^\circ,0.3^\circ,1^\circ,3^\circ\}$. Each continuous line represents a power, $\ell$, and the dashed line represents the term $\fnd$.}
    \label{fig:singlesourceangres}
\end{figure*}

In the ballistic regime, on the other hand, the dipolarity of the distribution is suppressed for large values of $\lsca /r_{s}$ as shown by the angular power spectrum. From equation~\ref{eq:ballistic}, it was expected that, in this regime, the angular distribution would be described by a delta function at $\theta = 0$. For such distribution, the amplitudes of $a_0$ and $a_1$ would tend to zero, while an infinite number of poles would be needed to describe it. A delta distribution, however, is non-physical since the source has a given size and the experiments have a given angular resolution. Figure~\ref{fig:singlesourceangres} shows the angular power spectrum when different angular resolutions, $\ares$, are tested. $C_{\ell}$ tend to a finite value which is smaller the better the angular resolution and which follows $C_0 = C_1/3$, as expected from the definition in equation~\ref{eq:coefficients}. Throughout the paper $\theta_{\mathrm{res}} = 1^\circ$ is used.

\section{First approximation and variance of the total dipole}
\label{app:approximation}

We consider a set of distance shells, each located at $i\Dmin$, and containing $i^2$ number of sources within the shell volume. The contribution of each shell to the final dipole can be estimated as

\begin{equation}
    \left \langle \Phi_1^{(i)}\right \rangle^2 = \frac{i^2 \left(\Phi_1^{(s_i)} n_{s_i}\right)^2}{\ntot^2},
\end{equation}

where $\ntot$ is total cosmic ray density coming from all the sources, and $a_1^{(s_i)}$ and $n_{s_i}$ are, respectively, the dipole term and the cosmic ray density from a single source in shell $i$. In the steady-state diffusion regime, the cosmic ray density from a single source in shell $i$ behaves as $1/r_i = 1/(i\Dmin)$ and the dipole strength from such a source behaves as $\lsca/r_i = \lsca/(i\Dmin)$. The overall contribution of sources in shell $i$ to the dipole, can therefore be as a function of the contribution of the first source,

\begin{equation}
    \left \langle \Phi_1^{(i)}\right \rangle^2 = \frac{i^2 \left(\frac{\Phi_1^{(1)} n_1}{i^2}\right)^2}{\ntot^2} = \left( \frac{\Phi_1^{(1)} n_1}{i \ntot}\right)^2.
\end{equation}

Consequently, the total dipole can be approximated as

\begin{equation}
\begin{split}
    \left \langle \Phi_1^{\mathrm{(tot)}}\right \rangle^{2} &\approx \sum_i \left( \frac{\Phi_1^{(1)} n_1}{i \ntot}\right)^2 \implies \\
    \delta = \left \langle \Phi_1^{\rm (tot)} \right \rangle &\approx \frac{\Phi_1^{(1)} n_1}{\ntot}.
\end{split}
\end{equation}

Throughout the paper, we have considered sources randomly distributed, leading to $\langle \cos (n\alpha) \rangle=0$ in equation~\ref{eq:cos}. Therefore the $\Phi_1$ term here calculated represents the mean dipole. It is important to estimate its variance.

We use equation~\ref{eq:cos} and consider sources in the same shell, i.e., with the same $\Phi^{(s)}_1$ term for the single source angular distribution. For the convenience of the notation, we define the dipole term of the summation of two angular distributions as (see equation~\ref{eq:cos})

\begin{equation}
A = \left(2 \Phi_1^{(\rm sum)}\right)^2.
\end{equation}

The variance will then be given by

\begin{equation}
\begin{split}
    \left\langle A \right\rangle &= 2 \Phi_1^2 + 2 \Phi_1^2 \langle\cos(\phi)\rangle = 2 \Phi_1^2 \\
    \left\langle A^2 \right\rangle &= \left(2 \Phi_1^2\right)^2 + 4 \Phi_1^4 \langle \cos^2(\phi) \rangle + 4 \Phi_1^4 \langle \cos (\phi) \rangle = \\
    &= \left(2 \Phi_1^2\right)^2 + 4 \Phi_1^4 \\
    \sigma \left(A\right) &= \sqrt{\langle A^2 \rangle - \langle A \rangle^2} = 2 \Phi_1^2,
    \end{split}
\end{equation}

from which, one can obtain the total variance,

\begin{equation}
\begin{split}
    \sigma\left(\Phi_1^{\rm tot}\right) = \sigma(\delta) = \sqrt{\sum_i \left(\frac{\Phi_1^{(i)}}{\sqrt{i}}\right)^2} \approx \\ \approx \sqrt{\sum_i \left(\frac{\Phi_1^{(1)}n_1}{i^2 \ntot}\right)^2} \approx \delta.
    \end{split}
\end{equation}

Due to the Poissonian behavior, the variance is dictated by the closest shells, which contain fewer sources and, as demonstrated, is of the same order of the mean. 

Another source of uncertainties in the model is the relation between the distance to the nearest source, $\Dmin$, and the average distance between sources, which is taken as the same for the hypothesis here considered. A large fluctuation is expected, though, for the distance to the nearest source, which is known to drive the dipole.

\end{appendix}

\end{document}